\documentclass[12pt]{article}
\usepackage{graphicx}
\usepackage{amsmath}
\usepackage{amsfonts}
\usepackage{amssymb}
\usepackage{verbatim}
\usepackage{hyperref}
\usepackage{cite}

\newcommand{\Letter}{
    \setlength{\textwidth}{7in}
    \setlength{\textheight}{9.5in}
    \hoffset=-0.75in
    \voffset=-1.15in }

\Letter

\renewcommand{\t}[1]{\tilde{#1}}

\newcommand{\be}{\begin{eqnarray}}
\newcommand{\ee}{\end{eqnarray}}
\newcommand{\kD}{\kappa_{D}^2}
\newcommand{\kp}{\kappa_{p+1}^2}
\newcommand{\next}{\nonumber \\}

\begin{document}

\begin{titlepage}
\setcounter{page}{1} \baselineskip=15.5pt \thispagestyle{empty}

\begin{center}
\mbox{}
\bigskip

{\LARGE A Breathing Mode for Warped Compactifications}
\end{center}
\bigskip

\begin{center}
Bret Underwood
\end{center}

\begin{center}
Department of Physics, McGill University, Montr\'eal, QC H3A 2T8 Canada \\ bret.underwood@mcgill.ca
\end{center}

\begin{center}
{\bf
Abstract}
\end{center}

\noindent In general warped compactifications, non-trivial backgrounds for the warp factor and the 
dilaton break $D$-dimensional diffeomorphism invariance, so that dilaton fluctuations can be gauged
away completely and eaten by the metric.  More specifically, the warped volume 
modulus and the dilaton are not independent, but combine into a single gauge-invariant degree of freedom in
the lower dimensional effective theory, the warped breathing mode.  
This occurs for all strengths of the warping, even the weakly warped limit.
This warped breathing mode appears as a natural zero mode deformation of backgrounds 
sourced by p-branes, and affects the identification of the independent degrees of freedom 
of flux compactifications.


\end{titlepage}


\section{Introduction}
\setcounter{equation}{0}

The effective theory describing the low energy limit of a compactification 
contains many light degrees of freedom, particularly
scalar fields, arising as fluctuations of the higher dimensional fields.  In string theory compactifications 
two such degrees of freedom are universal, appearing in any compactification: the volume modulus, 
a fluctuation of the metric
that controls the volume of the compact space, and the dilaton, a fluctuation of the $10$-dimensional
scalar field that controls the strength of the string coupling.  
While these two degrees of freedom, and their effective theory, are most easily analyzed in the unwarped limit, 
many compactifications of phenomenological interest contain matter and localized objects that source non-trivial
warping 
\cite{hep-th/9605053,hep-th/9908088,hep-th/0004103,KKLT,GKP,FreyPolchinski,hep-th/0308156,Douglas1,Douglas2}.  
Often, the way a lower dimensional degree of freedom 
appears in the higher dimensional fields 
in the presence of warping can look very different from its relatively simple unwarped form; for
some examples see \cite{GM, STUD, FTUD}.
Motivated by the simplicity and universality of the volume modulus and the dilaton, in this paper we
will construct their higher-dimensional origin in warped backgrounds.
We will show that in a warped background the (warped) volume modulus and the dilaton
combine into a single degree of freedom, which we will call the warped breathing mode.
As we will explain in more detail below,
these degrees of freedom combine because of the spontaneous breaking 
of the higher-dimensional diffeomorphisms by the
warping and the existence of non-trivial constraint equations arising from the higher-dimensional Einstein equations.

First, let us first review how fields, and their perturbations, behave on warped backgrounds.
The study of dynamics and fluctuations on warped compactifications is much more complicated than the standard
Kaluza-Klein (KK) reduction on direct product spaces, and has been studied extensively by a number
of authors 
\cite{GM, STUD, FTUD, vandeBruck, Csaki:2000zn, Tanaka:2000er, GravitationalHiggs, Rabadan:2002wy, KofmanWarpedPertTheory, DG, da1, buchel, da2, WarpedSUSY, dst,DT, FreyMaharana,Koerber:2007xk,Martucci:2009sf,NakayamaD3,WarpedOpen}.  
A number of well-known physical effects contribute to the challenges of warped effective theories:
\begin{itemize}
\item Wavefunctions in the internal space localize to regions of strong warping, so that terms in the effective
theory involving wavefunction overlaps are more difficult to compute.
\item In a KK reduction fields are expanded in a KK tower; usually only the zero mode is kept, and the higher modes
integrated out, because the zero mode mass is hierarchically smaller than the KK mode masses (set by the
KK scale).  In spaces with regions of strong warping, however, the masses of KK modes are strongly redshifted
to the IR, so that they cannot in principle be integrated out.
\end{itemize}
These effects are best understood through a simple model of a scalar field in a warped space \cite{FreyMaharana}.  
In particular,
consider the $D$-dimensional warped product of a $(p+1)$-dimensional (external) 
spacetime and a $(D-p-1)$-dimensional
(internal) compact space, spanned by $x^\mu, y^m$ respectively, with the background metric
\be
ds_D^2 = e^{2A_0(y)} \hat g_{\mu\nu} dx^\mu dx^\nu + g_{mn}(y) dy^m dy^n\, .
\label{eq:warpedintro}
\ee
The function $A_0(y)$ is known as the warp factor.  In addition to gravity, we will allow a $D$-dimensional
scalar field $\phi$ and $D$-dimensional matter, with the action
\be
S = \frac{1}{2\kappa_D^2} \int d^Dx \sqrt{g_D}\left[R_D - \frac{1}{2} (\partial \phi)^2 + {\mathcal L}_m\right]\, .
\label{eq:scalaraction}
\ee
We will refer to the $D$-dimensional scalar field $\phi$ as the {\it dilaton} field throughout the rest of the paper,
in obvious analogy with the dilaton of $10$-dimensional supergravity theories.
The rest of the matter in ${\mathcal L}_m$ will contribute to generating the warped background, but 
fluctuations in these fields will not be important, as we will see.

Fluctuations of the dilaton on the background (\ref{eq:warpedintro}) take the form
\be
\phi(x,y) = \phi_0(y) + \delta \phi(x,y) = \phi_0(y) + \sum_n \delta \phi_n(x) \t \phi_n(y),
\label{eq:dilatonKKintro}
\ee
where $\phi_0(y)$ is the background profile of the dilaton.  Since the dilaton typically also couples to the
$D$-dimensional matter generating the warp factor, we will refer to non-trivial 
background profiles for $A_0(y), \phi_0(y)$ as ``warping" in general\footnote{A few special cases
exist, for example the string theory GKP \cite{GKP} backgrounds with constant dilaton still have 
non-trivial metric warping.}.
In writing (\ref{eq:dilatonKKintro}) we expanded the fluctuation into a tower of KK degrees of freedom 
$\delta \phi_n(x)$.
The equation of motion for $\delta \phi(x,y)$, allowing for a $D$-dimensional mass $m_\phi^2$ (which can
arise e.g.~from fluxes, see \cite{FreyMaharana}), is
\be
 \nabla_D^2\delta \phi(x,y) - m_\phi^2 \delta \phi(x,y) &=& \sum_n \left[e^{-2A_0(y)} \t \phi_n(y) \hat \Box \delta \phi_n(x) + (p+1) (\partial^p A_0)(\partial_p \t \phi_n(y)) \delta \phi_n(x)\right. \next
&& \left. + \nabla_{D-p-1}^2 \t \phi_n(y) \delta \phi_n(x) - m_\phi^2 \t \phi_n(y) \delta \phi_n(x)\right] = 0.
\ee
Writing this in terms of the $(p+1)$-dimensional mass $\hat \Box \delta \phi_n(x) = m_n^2 \delta \phi_n(x)$,
we have a Schr\"odinger-type equation for the wavefunction $\t \phi_n(y)$ on the compact internal space:
\be
\nabla^2_{D-p-1} \t \phi_n(y) + (p+1)(\partial^p A_0)(\partial_p \t\phi_n) + \left(e^{-2A_0}m_n^2 -m_\phi^2\right) \t \phi_n(y) = 0 .
\ee
The naive zero mode $\delta \phi_0(x) \t \phi_0$ is usually taken to be constant,
with a $(p+1)$-dimensional mass $m_0 \sim m_\phi$.
This may not, however, be the lowest mass mode of the background.
As discussed in \cite{FreyMaharana}, the wavefunction for KK modes localizes to regions of strong warping
(called warped throats) where $e^{-2A_0} \gg 1$.  The warped throat acts like a gravitational well,
redshifting the masses of the lowest KK modes to
$m_n \sim e^{A_0} m_\phi \ll m_0 \nonumber$
so that the KK modes are much lighter than the naive zero mode.  Thus, they cannot be integrated out, and must
be included in the low energy effective field theory.  Similar conclusions result when examining perturbations
of the metric \cite{GM,STUD,FTUD,DT}.

The above warping effects arise just by examining the dilaton equation of motion, which treats the dilaton as
a probe field in a fixed warped background.  When the dilaton is coupled to gravity, however, additional interesting
effects due to warping arise:
\begin{itemize}
\item Diffeomorphisms mix fields in completely different sectors (e.g.~gravity and dilaton fields).  Only
combinations of fields from different sectors are true {\it gauge-invariant} degrees of freedom.
\item Constraint equations arising from the higher dimensional equations of motion constrain fields to be dependent
on each other in a non-dynamical way, reducing the overall number of {\it independent} degrees of freedom.
\end{itemize}

In particular, consider a generic fluctuation of the dilaton
$\phi(x,y) = \phi_0(y) + \delta \phi(x,y)$.  
When we allow the metric to be dynamical, it is possible to transfer the entire dilaton
fluctuation into the metric through a $D$-dimensional diffeomorphism 
$\xi^M(x,y)$ since
\be
g_{MN} &\rightarrow & g_{MN} - \nabla_M \xi_N - \nabla_N \xi_M; \next
\phi(x,y) &\rightarrow & \phi_0(y) + \delta \phi(x) \t\phi(y) + \xi^m \partial_m \phi_0 = \phi_0(y); \next
&&\mbox{with\ } \xi^m(x,y) = -\frac{\partial^m \phi_0(y)}{(\partial \phi_0)^2} \delta \phi(x,y).
\ee
Because of the background profile $\phi_0(y)$, the dilaton fluctuation is no longer gauge-invariant,
so it does not make sense to talk about it as a separate degree of freedom from the metric.
The phenomenon arises because the background profile $\phi_0(y)$ spontaneously breaks the $D$-dimensional
diffeomorphism invariance into a preferred slicing of $(p+1)$- and $(D-p-1)$-dimensional spaces.
In an analogous way to spontaneously broken gauge theories, the dilaton fluctuation $\delta \phi(x,y)$
can then be eaten by the metric in a suitably chosen (unitary) gauge (see \cite{Creminelli:2006xe, Cheung:2007st}
for more discussion on the analogy with spontaneously broken gauge theories in the context of cosmology).

Next, we note that a pure dilaton fluctuation gives rise to an off-diagonal $(\mu m)$ component of the $D$-dimensional
Einstein equations:
\be
0=\delta G_{\mu m} - \kappa_D^2 \delta T_{\mu m} = 
	-\frac{1}{2} \left(\partial_\mu \delta \phi(x)\right) \t \phi(y) \partial_m \phi_0(y).
\label{eq:scalarmumintro}
\ee
If no other fluctuations are turned on, this equation implies that the dilaton cannot be dynamical 
$\partial_\mu \delta \phi(x) = 0$.  However, inspired by our previous observation about dilaton-metric mixing
through diffeomorphisms, metric fluctuations should also be included; the metric fluctuations will contribute additional
terms to (\ref{eq:scalarmumintro}) so that it can be consistently solved.
In particular, turning on a metric fluctuation $\delta g_{mn} = u(x) \delta_u g_{mn}(y)$, (\ref{eq:scalarmumintro})
becomes (schematically):
\be
\delta G_{\mu m} - \kappa_D^2 \delta T_{\mu m}  \sim \left(\partial_\mu u(x)\right) \nabla^n \delta_u g_{mn} 
	- \frac{1}{2}\left(\partial_\mu \delta \phi(x)\right) \t \phi(y) \partial_m \phi_0(y) = 0\, .
\label{eq:mumintro}
\ee
This can only be solved if the metric and dilaton degrees of freedom are identified with each other 
$u(x) \sim \delta \phi(x)$.
More precisely, metric and dilaton fluctuations can be written in gauge invariant combinations, and these gauge
invariant fluctuations are coupled through the Einstein equations.

These off-diagonal components of the Einstein equations are just the usual momentum constraint equations of the
Hamiltonian formulation of general relativity (see \cite{Wald}).  The existence of a non-zero constraint equation
is related to the fact that the dilaton fluctuation transforms under diffeomorphisms: as is typical in gauge
theories, constraint equations act as the generators for gauge transformations.  Thus, the two effects
of gauge (diffeomorphism) non-invariance and non-zero constraint equations are just two sides of the same
phenomenon.

It is important to appreciate that the mixing of the dilaton and metric degrees of freedom through the two
effects just discussed is non-dynamical, in the sense that the dilaton degree of freedom $\delta \phi$ 
is not acting as a source for the metric degrees of freedom $\delta g$,
but rather must be identified with the metric degrees of freedom.
In particular, the off-diagonal Einstein equations (\ref{eq:mumintro}) arise as initial
value constraints on the configuration space, not as dynamical equations of motion: 
since they do not involve second order time derivatives, they must be imposed for all time.

Similar effects happen in $(3+1)$-dimensional cosmological backgrounds with a homogeneous time-dependent
scalar field: diffeomorphisms mix the scalar field and metric perturbations, so that only gauge invariant combinations
of these fields are physical, and the constraint equations couple the gauge-invariant metric and scalar field fluctuations.
Pertubations can be studied using the well-developed formalism of cosmological perturbation theory \cite{Bardeen:1980kt,Mukhanov:1990me}, which
emphasizes the construction of gauge-invariant variables and the role of constraints.
One interesting result from this formalism is that there exists a gauge (comoving gauge) in which the scalar
degree of freedom $\zeta$ is encoded as a volume rescaling factor of the spatial metric
$g_{ij} = a^2(t) e^{2\zeta} \delta_{ij}$.

A similar formalism can be developed for perturbations in warped backgrounds 
\cite{vandeBruck,KofmanWarpedPertTheory}.
The relevant metric fluctuations are the scalar components
(with respect to the $(p+1)$-dimensional spacetime) of the $D$-dimensional graviton, and correspond
to deformation modes of the internal space.  
This presents a puzzle: which deformation mode of the internal metric should the dilaton mix with through 
diffeomorphisms and the constraint equations?  Fortunately, intuition from the cosmological case suggests the
answer: the dilaton should mix with the ``volume rescaling factor" of the internal space, otherwise known as
the warped volume modulus.
Mixing between a bulk scalar field
and the volume modulus is warped backgrounds has been seen previously for Randall-Sundrum (RS)
models \cite{RS1,RS2}.  There, fluctuations in the bulk scalar field used in
the Goldberger-Wise stabilization mechanism for the radion \cite{GW} are coupled to 
fluctuations in the radion itself \cite{Csaki:2000zn,Tanaka:2000er, GravitationalHiggs}
(for more recent investigations, see \cite{George1, George2}).

As a specific example, consider the $10$-dimensional supergravity limit of string theory with fluxes and localized
sources such as O-planes and D-branes, where the dilaton is the usual
string theory dilaton (in $10$-dimensional Einstein frame).
The fluxes and localized objects act as sources for the dilaton and warp factor,
so that both the dilaton and warp factor have non-trivial profiles on the internal space\footnote{Again, 
GKP \cite{GKP} backgrounds where the dilaton is constant are an exception.}.  
A priori it would seem natural
to regard fluctuations of the dilaton and the volume modulus as two independent degrees of freedom.
However, as argued above, and will be shown in more detail in the rest of the paper, the behavior of fluctuations 
under diffeomorphisms and the existence of constraint equations imply this cannot be the case: 
the fluctuations of the volume modulus and dilaton are controlled by a single degree of freedom,
the {\it breathing mode}, which has as its $10$-dimensional
wavefunction a mixture of the warped volume modulus and the dilaton\footnote{This
is different than the kinetic mixing between the dilaton and the volume modulus that arises when working
in $10$-dimensional string frame.  This latter effect disappears after a field redefinition, which is equivalent
to working in $10$-dimensional Einstein frame, and preserves the number of degrees of freedom.  In contrast, the
mixing we are pointing out occurs even in Einstein frame, and changes the number of degrees of freedom.}.

In Section \ref{sec:PertTheory}, 
we review the formalism of cosmological perturbation theory \cite{Bardeen:1980kt,Mukhanov:1990me},
and its application to perturbations on warped compactifications 
\cite{vandeBruck,KofmanWarpedPertTheory}.
The formalism of warped perturbation theory, namely the construction of gauge-invariant variables and
the role of the constraint equations, illustrates the non-dynamical mixing between the dilaton and the metric.
In the rest of the paper we will illustrate how in warped backgrounds mixing occurs between the
warped volume modulus and the dilaton.
In Section \ref{sec:BreathingMode} we show how 
the dilaton and warped volume modulus combine 
through the constraint equations, and compute the kinetic term of the resulting ``breathing mode"
degree of freedom in the dimensionally reduced effective theory.  
In Section \ref{sec:pbranes} we show how the breathing mode also
appears naturally as a zero mode of p-brane-like compactifications.  
We conclude with a discussion of the implications of our results in Section \ref{sec:Discussion}.
Appendix \ref{sec:WarpedVolMod} contains
the metric wavefunction for the warped volume modulus for arbitrary spacetime dimensions.
Throughout the paper, the term ``degree of freeedom" always refers to a $(p+1)$-dimensional field,
while the term ``field" is reserved for $D$-dimensional fields.

\section{Cosmological Perturbation Theory and Warped Compactifications}
\label{sec:PertTheory}
\setcounter{equation}{0}

It was noted in \cite{vandeBruck,KofmanWarpedPertTheory} that there is a similarity 
in the structure of perturbations on cosmological and
warped backgrounds.  These similarities are useful for understanding the mixing of warped degrees of freedom
through diffeomorphisms and constraints that we will focus on throughout the paper.  In this section,
we will first review cosmological perturbation theory for a scalar field on a cosmological background.
We will then review the application of this formalism to warped perturbation theory for perturbations of
the dilaton and the metric.
In later sections, we will specialize to the case where the metric perturbations correspond to the warped
volume modulus.

\subsection{Cosmological Perturbation Theory}
\label{sec:CosmoPertTheory}

We will follow the general formalism for cosmological perturbation theory of \cite{Mukhanov:1990me}.
Our background is a 4-dimensional FLRW spacetime on flat 3-dimensional space:
\be
ds^2 = -dt^2 + a^2(t) \delta_{ij} dx^i dx^j,
\ee
with scalar field matter that has a homogeneous background profile $\phi = \phi(t)$.  Scalar perturbations (with
respect to the spatial directions $\vec{x}$) about
this background take the form,
\be
ds^2 &=& -(1+2\varphi(t,x)) dt^2 + a^2(t) \left[(1-2\psi(t,x))\delta_{ij} + 2 \partial_i \partial_j E(t,x)\right] dx^i dx^j \next
	&&+ a(t) \partial_i B(t,x) dt dx^i; \\
\phi &=& \phi_0(t) + \delta\phi(t,x)\, .
\ee
Altogether we have 5 scalar fluctuations $\{\varphi, \psi, E,B,\delta\phi\}$.  However, not all of these fluctuations
are independent.  Under infinitesimal coordinate transformations 
$x^\mu \rightarrow x^\mu + \xi^\mu(t,x)$
the metric and scalar field transform as
\be
g_{\mu\nu} &\rightarrow & g_{\mu\nu} - 2 \nabla_{(\mu} \xi_{\nu)}; \\
\phi &\rightarrow & \phi + \xi^\mu \partial_\mu \phi\, .
\ee
In particular, spatial {\it scalar} diffeomorphisms $\xi^\mu = \{\xi^0(t,x),\delta^{ij}\partial_j \lambda(t,x)\}$ can
be used to shuffle degrees of freedom between the metric and scalar matter sectors:
\be
\varphi &\rightarrow & \varphi - \dot \xi^0; \next
\psi &\rightarrow & \psi + \frac{\dot a}{a} \xi^0; \next
B &\rightarrow & B + \frac{1}{a} \xi^0 - a \dot \lambda ; \next
E &\rightarrow & E - \lambda ; \next
\delta \phi &\rightarrow & \delta \phi + \dot \phi_0 \xi^0;
\ee
(where a dot denotes a derivative with respect to time $t$).
Clearly, the original scalar fluctuation variables are not gauge invariant
but gauge invariant variables can be constructed \cite{Bardeen:1980kt,Mukhanov:1990me}:
\be
\Phi_B &=& \varphi - \frac{d}{dt} \left[a^2 (\dot E - B/a)\right]; \\
\Psi_B &=& \psi + \frac{\dot a}{a} a^2 (\dot E-B/a); \\
\delta \Phi &=& \delta \phi + a^2 \dot \phi_0 (\dot E - B/a)\, .
\ee 
Of the 5 original scalar fluctuations, only 3 of them are gauge invariant; 2 of the scalar fluctuations are gauge
artifacts, and can be removed by an appropriate gauge transformation.

Let us now examine the constraint equations arising from the Einstein equations.
The standard energy and momentum constraint equations arise as the time-time and time-space components
of the Einstein equations.  Written in terms of the gauge invariant variables above, they are (to
first order in the fluctuations):
\be
\delta G_{00} - 8\pi G \delta T_{00} &=& 3 H \left(\dot \Psi_B+H \Phi_B\right) + \nabla^2 \Psi_B + 4\pi G \delta \rho = 0 ;\\
\delta G_{0i} - 8\pi G \delta T_{0i} &=& 2 \partial_i \left[\dot \Psi_B + \frac{\dot a}{a} \Phi_B - 4\pi G \dot \phi_0 \delta \Phi\right] = 0\, .
\ee
where $\delta \rho = \frac{1}{2} \dot{\delta \Phi} \dot \phi_0 + \delta \Phi V'(\phi_0)$.
These equations do not contain second order time derivatives, so are not dynamical equations of motion.
Instead, they are initial value constraints that must be imposed for all time.
Imposing these constraint equations reduces the total number of independent degrees of freedom.
As there are two independent constraint equations for our 3 gauge-invariant fields $\Phi_B,\Psi_B,\delta \Phi$,
we are left with only a single independent gauge-invariant degree of freedom.

\subsection{Warped Perturbation Theory}
\label{sec:warpedperttheory}

Inspired by the similarities between perturbations in cosmological and warped backgrounds, 
in this section we will develop a similar formalism for warped perturbation theory, following
\cite{vandeBruck,KofmanWarpedPertTheory}.

We will take a $D$-dimensional warped product of a $(p+1)$-dimensional external spacetime and a
$(D-p-1)$-dimensional internal compact space, with the background,
\be
ds_D^2 &=& e^{2A_0(y)} \hat g_{\mu\nu}(x) dx^\mu dx^\nu + e^{-2B_0(y)} \t g_{mn}(y) dy^m dy^n; \next
\phi &=& \phi_0(y)\, .
\label{eq:warpedBackground}
\ee
Greek indices run over the external spacetime $\mu,\nu = 0...p$ while lower-case latin indices run over
the internal space $m,n = p+1...D-1$.  
We are assuming that there are other matter fields with Lagrangian ${\mathcal L}_m$ as well, 
giving rise to a background energy-momentum tensor
$T_{MN}^{(0)} = \left\{T_{\mu\nu}^{(0)},T_{mn}^{(0)}\right\}$ 
so that the background (\ref{eq:warpedBackground}) is a solution to the background equations
of motion.
In particular, we have in mind a p-brane like background, which is discussed in more detail in Section \ref{sec:pbranes}.
We will take the background external spacetime to be some $(p+1)$-dimensional maximally symmetric space 
$\hat g_{\mu\nu}$ such as anti-de Sitter (AdS), Minkowski,
or de Sitter (dS) space.  We will leave
the compact internal metric $\t g_{mn}$ arbitrary, to the extent that it is a background solution; 
for example, the curvature of the internal space may be positive, negative,
or zero, according to the solution to the background Einstein equations with the choice of matter.
The warp factor $e^{-2B_0(y)}$ is pulled out of the internal metric by convention, and will be chosen to be
$B_0(y) = (p+1)/(D-p-3) A_0(y)$, again by convention (occasionally we will keep expressions in terms
of $B_0$ for compactness).

Perturbations in the metric and dilaton\footnote{We will restrict perturbations in the matter sector to the dilaton in
this paper.  This represents a truncation of the most general set of perturbations; we leave the study of more
general perturbation ans\"atze to future work.}
about this background that are scalar with respect to the $(p+1)$-dimensional spacetime take
the form:
\be
\label{eq:warpedpertmetric}
ds_D^2 &=& e^{2A_0(y)} \left[ (1-2 \psi(x,y)) \hat g_{\mu\nu} + 2 \hat \nabla_\mu \partial_\nu E(x,y)\right] dx^\mu dx^\nu + e^{2 A_0(y)} \partial_\mu K_m(x,y) dx^\mu dy^m \next
&& + e^{-2B_0(y)} \left[\t g_{mn}(y) + 2 \varphi_{mn}(x,y)\right] dy^m dy^n; \\
\phi &=& \phi_0(y) + \delta \phi(x,y)\, .
\label{eq:warpedpertdilaton}
\ee
As in the cosmological case, not all of these scalar fluctuations $\{\psi,E,K_m,\varphi_{mn},\delta\phi\}$
are independent because of their behavior under $D$-dimensional diffeomorphisms
\be
\begin{pmatrix}
x^\mu \\
y^m \\
\end{pmatrix}
\rightarrow
\begin{pmatrix}
x^\mu + \xi^\mu(x,y) \\
y^m + \xi^m(x,y)\\
\end{pmatrix}
\, . \label{eq:diff}
\ee
The relevant $D$-dimensional diffeomorphisms are those which behave as scalars with respect to
the $(p+1)$-dimensional spacetime: $\xi^M = \{\hat g^{\mu\nu} \partial_\nu \hat \lambda(x,y),\xi^m(x,y)\}$.
The scalar fluctuations transform under these diffeomorphisms as
\be
\label{eq:varphigauge}
\varphi_{mn} &\rightarrow & \varphi_{mn} - \t \nabla_{(m} \xi_{n)} + \left(\xi^p \partial_p B_0\right) \t g_{mn}; \\
\psi &\rightarrow & \psi + \xi^m \partial_m A_0;  \\
E &\rightarrow & E - \hat \lambda ; \\
K_m &\rightarrow & K_m - \partial_m \hat \lambda - e^{-2B_0-2A_0} \t g_{mn} \xi^n ; \\
\delta \phi &\rightarrow &  \delta \phi + \xi^m \partial_m \phi_0 ;
\label{eq:dilatongauge}
\ee
where $\hat g_{\mu\nu}, \t g_{mn}$ always represent the background metric.
Here it is clear that the dilaton fluctuations $\delta \phi$ mix with the scalar metric fluctuations through
gauge transformations due to the background profile $\phi_0(y)$.  
This can be understood in the language of spontaneously broken gauge theories:
the background profile $\phi_0(y)$ spontaneously breaks the $D$-dimensional diffeomorphism (gauge) invariance,
and the $D$-dimensional graviton (gauge field) can gain an extra degree of freedom by eating 
the dilaton fluctuation.

We can again define a set of gauge-invariant fields (indices on partial derivatives are raised with the warped
metric $g^{mn} = e^{2B_0} \t g^{mn}$)
\be
\label{eq:GaugeInvPhi}
\Phi_{mn} &=& \varphi_{mn} + e^{2A_0} (\partial^p B_0)(K_p-\partial_p E) \t g_{mn}
	+ \t\nabla_{(m} \left[e^{2A_0+2B_0}(\partial_{n)} E - K_{n)})\right] ; \\
\label{eq:GaugeInvPsi}
\Psi &=& \psi + e^{2A_0} (\partial^p A_0)(K_p - \partial_p E); \\
\label{eq:GaugeInvDPhi}
\delta \Phi &=& \delta \phi + e^{2A_0}(\partial^p \phi_0)(K_p - \partial_p E)\,.
\ee
The diffeomorphism transformations have removed $(D-p)$ of the original scalar fluctuations.
Eq.(\ref{eq:GaugeInvDPhi}) shows how the scalar metric fluctuations mix with the dilaton fluctuations
to create the gauge-invariant dilaton fluctuation $\delta \Phi$ when the dilaton has a non-trivial
background profile.

The gauge-invariant variables also must satisfy constraints coming from the Einstein equations.
For perturbations about warped backgrounds, these constraints arise from the off-diagonal 
$\mu \neq \nu$ and $\mu m$ Einstein equations\footnote{See \cite{STUD,DT} for more discussion
about the role of constraints in warped compactifications.}, which read 
(with $\Phi^{\t p}_p \equiv \Phi_{pq}\t g^{pq}$):
\be
\label{eq:GaugeInvConstraintmunu}
\left.\delta G_{\mu\nu} - \kappa_D^2 \delta T_{\mu\nu}\right|_{\mu\neq \nu} &=& 
	\hat\nabla_\mu \partial_\nu \left[(p-1)\Psi - \Phi_p^{\t p}\right]=0;  \\
\delta G_{\mu m} - \kappa_D^2 \delta T_{\mu m} &=& -\partial_\mu \partial_m \left[p \Psi + \Phi_p^{\t p}\right]
	+ \partial_\mu \t\nabla_p \Phi_m^{\t p} + \partial_\mu \Phi_p^{\t p} \left[\partial_m A_0 + \partial_m B_0\right] \next
	&& + \partial_\mu \Phi_m^{\t p} \left[(p-1)\partial_p A_0 - (D-p-1)\partial_p B_0\right] 
	+\frac{1}{2} \partial_\mu \delta \Phi \partial_m \phi_0 = 0\, . \hspace{.2in}
\label{eq:GaugeInvConstraintmum}
\ee
Notice that the dilaton fluctuation does not contribute at linear order to the $\mu\neq \nu$ constraint
equations, but it does contribute to the $\mu m$ constraint equations when the background profile
is non-trivial.  Thus, the gauge-invariant dilaton and metric fluctuations cannot
be independent.
For example, it is not possible to consider fluctuations of the dilaton by itself, since the constraint equation
(\ref{eq:GaugeInvConstraintmum}) is in that case
\be
\nonumber
\delta G_{\mu m} - \kappa_D^2 \delta T_{\mu m} = -\frac{1}{2} \partial_\mu \delta \phi \partial_m \phi_0 = 0
\ee
which cannot be solved for a dynamical dilaton fluctuation $\delta \phi$.

Altogether, we started with $3+(D-p-1)(D-p+2)/2$ scalar fluctuations; after applying $2(D-p)$ constraints
and gauge fixings, we are left with $1+(D-p-1)(D-p-2)/2$ independent and gauge-invariant scalar
fluctuations.  Note that this is the number of independent $D$-dimensional dilaton and metric 
fluctuations that are scalars in $(p+1)$-dimensions,
not the number of $(p+1)$-dimensional degrees of freedom.
The number of $(p+1)$-dimensional degrees of freedom is in fact infinite, consisting of an
infinite tower of KK-modes.
In particular, the $D$-dimensional fluctuations
$\{\Psi(x,y),\Phi_{mn}(x,y),\delta \Phi(x,y)\}$ 
can be expanded in appropriate eigenmodes of the warped internal
space, e.g.~$\Psi(x,y) = \sum_n u_n(x) Y^n(y)$ (see \cite{STUD} for
more discussion of eigenmodes for warped spaces), corresponding to an infinite tower of 
gauge-invariant, $(p+1)$-dimensional degrees of freedom.

This approach of identifying gauge-invariant variables and coupling them through the Einstein constraint equations
is equivalent to the Hamiltonian construction of warped perturbation theory \cite{DT,NakayamaD3}.
In the Hamiltonian formalism, one performs an ADM decomposition of the metric and constructs the canonical
momentum associated to the spatial metric, so that the Hamiltonian is written in terms of the canonical momentum.
Invariance of the Hamiltonian under gauge transformations then enforces the constraints.
While the Hamiltonian formalism is much more elegant, the slicing of spacetime into time and $(D-1)$-dimensional
space obscures the physical role the background dilaton profile $\phi_0(y)$ plays in inducing the preferred slicing into 
$(p+1)$- and $(D-p-1)$-dimensional spaces, which leads to the mixing.
Since it is this latter slicing that is most important for warped backgrounds, our approach is conceptually more
transparent for seeing the mixing between the dilaton and the volume modulus.

\section{Warped Breathing Mode}
\label{sec:BreathingMode}
\setcounter{equation}{0}

In the previous section we examined the formalism of warped perturbation theory for general scalar metric and dilaton
fluctuations (\ref{eq:warpedpertmetric},\ref{eq:warpedpertdilaton}) on the warped background 
(\ref{eq:warpedBackground}).  It was seen there that dilaton fluctuations generically mix non-dynamically 
with the metric fluctuations through the gauge-invariant fluctuations 
(\ref{eq:GaugeInvPhi}-\ref{eq:GaugeInvDPhi}) and the constraint equation (\ref{eq:GaugeInvConstraintmum}).
Only when the dilaton background is completely constant do the fluctuations in the different sectors decouple.

To illustrate this more explicitly, let us now restrict ourselves to a simple subset of fluctuations to show
how fluctuations from the metric and dilaton sectors combine into a single $(p+1)$-dimensional degree of
freedom.  We will call this degree of freedom the {\it warped breathing mode}, and it is composed of the
warped volume modulus and the dilaton fluctuation, which we will consider in turn.
To be clear, in this section we are considering one particular ansatz which consistently solves
the constraint equations; we are not providing the most general solution for perturbations in a warped background,
which is beyond the scope of this work.

\subsection{Ansatz}
\label{sec:WarpedBreathingAnsatz}

Let us begin with the metric sector, and consider the warped volume modulus 
in a general $D$-dimensional space.  This will be a generalization of the warped volume
modulus for compactifications from $10$ to $4$ dimensions given in \cite{FTUD}.
More details on the construction of the warped volume modulus can be found in Appendix \ref{sec:WarpedVolMod}.
As we saw in Section \ref{sec:warpedperttheory}, the most general form of scalar metric fluctuations
is given in (\ref{eq:warpedpertmetric}).  We will take these fluctuations to depend on a single $(p+1)$-dimensional
degree of freedom $u(x)$.  In the absence of dilaton fluctuations, $u(x)$ would be identified as the
warped volume modulus (as seen in Appendix \ref{sec:WarpedVolMod}).
The $y$-dependent parts of the metric fluctuations will be referred to as the metric wavefunction; e.g.~the
fluctuations can be expanded as $\psi(x,y) = u(x) \t \psi(y), \varphi_{mn}(x,y) = u(x) \t \varphi_{mn}(y)$, etc.

In order to differentiate the warped volume modulus from other possible metric deformation modes we will require
that it satisfy a few simple properties.
First, the fluctuation should correspond (for some fixed gauge) to a pure trace fluctuation of the internal metric
$\varphi_{mn} \propto \t g_{mn}$.  This is motivated by the form of the unwarped volume modulus 
(\ref{eq:unwarpedmetric}).  In the warped case, however, we expect that this fluctuation may obtain some non-trivial
wavefunction $\varphi_{mn} = u(x) \t\varphi(y)\t g_{mn}$ due to the background warping.
Note that while it seems natural to take $\varphi_{mn}$ to be pure trace, this is not a gauge-invariant statement.
The full gauge-invariant internal metric fluctuation $\Phi_{mn}$ (\ref{eq:GaugeInvPhi}) will not in general
be pure trace, even if $\varphi_{mn}$ is.
Since $\varphi_{mn}$ and $\psi$ are coupled through the constraint equations 
(\ref{eq:GaugeInvConstraintmunu},\ref{eq:GaugeInvConstraintmum}), we will also
need to turn on the fluctuation $\psi = u(x) \t \psi(y)$.  In general, the constraint equations also require
non-zero $K_m,E$ fluctuations, but since $\varphi_{mn}$ is pure trace, $K_m$ is not sourced through
the constraint equations or equations of motion, so we can set $K_m = 0$ without loss of generality\footnote{More
precisely, $K_m = \partial_m K$, and there always exists a gauge where this can be shifted into the metric
fluctuation $E$.}.

Next, we want the breathing mode to correspond to a fluctuation in the warped volume.  
The relevant warped volume appears in the dimensional reduction of the $D$-dimensional Ricci scalar
(we will suppress subscripts on metric determinants as 
$\mbox{det}\, \hat g_{\mu\nu} = \hat g,\ \mbox{det}\,\t g_{mn} = \t g$):
\be
\frac{1}{2\kappa_D^2}\int \sqrt{g_D}\, R_D \supset \frac{1}{2\kappa_D^2} \int \sqrt{\hat g}\, \hat R_{p+1} \int \sqrt{\t g}\, e^{-(p-1)A_0-(D-p-1)B_0} = \frac{1}{2\kappa_D^2} \int \sqrt{\hat g}\,\hat R_{p+1} \t V_W^{(0)}\ \ 
\ee
where
\be
\t V_W^{(0)} \equiv \int \sqrt{\t g}\, e^{-(p-1)A_0 - (D-p-1)B_0} = \int \sqrt{\t g}\, e^{-2\gamma A_0}
\ee
is the background warped volume, and $\gamma \equiv (D-2)/(D-p-3)$.
Finally, we would like the metric fluctuations to reduce to the unwarped volume modulus (reviewed in Appendix
\ref{sec:unwarped}) in the unwarped limit.

In particular, we will take our ansatz for the warped volume modulus deformation of the metric to be:
\be
ds^2 &=& e^{2A(y,u(x))} e^{2\Omega[u(x)]} \left[\hat g_{\mu\nu} + 2 e^{(p-3)\Omega[u(x)]}\hat \nabla_\mu \partial_\nu u(x) E(y)\right] dx^\mu dx^\nu , \next
&& + e^{-2\left(\frac{p+1}{D-p-3}\right) A(y,u(x))} \t g_{mn}(y) dy^m dy^n
\label{eq:warpedvolmetric}
\ee
where we have promoted the warp factor to be a function of the warped volume modulus so that
at linear order in $u(x)$
\be
A(y,u(x)) &\approx & A_0(y) + u(x) \delta A(y) + {\mathcal O}(u^2),
\ee
for some $\delta A$ to be determined by solving the constraint equations.
We have also included a Weyl factor, defined as
\be
e^{(p-1)\Omega[u(x)]} = \frac{\int \sqrt{\t g}}{\int \sqrt{\t g}\ e^{-2\gamma A(y,u(x))}} 
 = \frac{\t V_{D-p-1}}{\t V_W} ,
\label{eq:Weyl}
\ee
so that the dimensionally reduced Ricci scalar is in Einstein frame, where again to linear order
$\Omega[u(x)] \approx \Omega_0 + u(x) \delta \Omega + {\mathcal O}(u^2)$ with
$e^{(p-1)\Omega_0} \equiv \t V_{D-p-1}/\t V_W^{(0)}$.  This implies $\kappa_D^2 = \t V_{D-p-1} \kappa_{p+1}^2$,
as in the unwarped case.  

Comparing (\ref{eq:warpedvolmetric}) to the general form for metric perturbations
(\ref{eq:warpedpertmetric}), we have the identifications:
\be \label{eq:warpedbreathansatzpsi}
\psi(x,y) &=& - u(x) \left(\delta A(y) + \delta \Omega\right); \\
E(x,y) &=& u(x) E(y); \\
K_m(x,y) &=& 0;\\
\varphi_{mn}(x,y)&=& -\frac{p+1}{D-p-3} u(x) \delta A(y) \t g_{mn}\, .
\label{eq:warpedbreathansatzvarphi}
\ee

Now we will consider the dilaton sector.  As in (\ref{eq:warpedBackground}) we will take the dilaton to have
some non-trivial background profile $\phi_0(y)$.  Let us write fluctuations of the dilaton as
\be
\phi(x,y) = \phi_0(y) + \delta \phi(x) \t \phi(y) ,
\label{eq:dilatonfluct}
\ee
where $\delta \phi(x)$ is the $(p+1)$-dimensional degree of freedom associated with the dilaton, and
the dilaton fluctuation obtains a non-trivial wavefunction $\t \phi(y)$ due to the warping.
We are writing the dilaton degree of freedom $\delta \phi(x)$ as a separate degree of freedom from the warped volume
modulus $u(x)$, but we know
that it cannot really be dynamically independent since gauge transformations can be used to shift
$\delta \phi(x,y) = \delta \phi(x)\t \phi(y)$ entirely into the metric.
In particular, the gauge invariant dilaton fluctuation (\ref{eq:GaugeInvDPhi}):
\be
\delta \Phi = \delta \phi(x) \t \phi(y) - u(x) e^{2A_0} (\partial^p \phi_0)(\partial_p E(y))
\ee
does not make sense unless $\delta \phi(x) \propto u(x)$.  We can absorb the proportionality constant
into the dilaton wavefunction $\t \phi(y)$ so that the degrees of freedom are identified with each other
$\delta \phi(x) = u(x)$.  We will call this degree of freedom, 
which sources both the $D$-dimensional warped volume
modulus and the $D$-dimensional dilaton fluctuations, the warped breathing mode.
It is clear that (\ref{eq:warpedbreathansatzpsi}-\ref{eq:dilatonfluct}) fix the gauge completely,
since it is not possible to make a gauge transformation of the form 
(\ref{eq:varphigauge}-\ref{eq:dilatongauge}) that preserves 
(\ref{eq:warpedbreathansatzpsi}-\ref{eq:dilatonfluct}).

Even without noticing that gauge transformations mix the dilaton and warped volume modulus
fluctuations together, the $(\mu m)$ constraint equation couples them in an unavoidable way:
\be
\delta G_{\mu m} - \kappa_D^2 \delta T_{\mu m} &=& (\partial_\mu u(x))\left(\frac{D-2}{D-p-3}\right)
	\left[\partial_m \delta A - (p+1) \delta A \partial_m A_0\right] \next
	&&- \frac{1}{2} (\partial_\mu \delta \phi(x)) \t \phi(y) \partial_m \phi_0 = 0\,.
\label{eq:dilatonConst}
\ee
Clearly, we cannot solve this constraint unless the warped volume modulus and dilaton fluctuations 
are related, so we are again led to $\delta \phi(x) \propto u(x)$.  Then $\t \phi(y)$
and $\delta A(y)$ are related through this constraint.
Note that it
is not possible to solve this constraint equation by introducing only an off-diagonal vector
metric component $K_m \neq \partial_m K$ instead\footnote{Recall that a total derivative vector compensator
$K_m = \partial_m K$ can always be shifted into the compensator $E(y)$ through a $(p+1)$-dimensional
gauge transformation that leaves (\ref{eq:dilatonConst}) unchanged.} 
of a warp factor fluctuation because the resulting constraint equation is inconsistent.  One way to see
this is to notice that the fluctuation of the dilaton $\delta \phi(x,y)$ is a scalar with respect
to the internal space, while the off-diagonal metric component $K_m$ is a vector.  Since vectors and
scalars transform differently under rotations of the internal space, it is not possible for
these types of terms to cancel in the constraint equations.  Alternatively, it is straightforward to compute the constraint
equation including only $K_m$, and it is seen to be inconsistent.
This conclusion about mixing between the volume modulus and dilaton fluctuations is quite general, since
the mixing seen in (\ref{eq:dilatonConst}) follows directly from the general metric ansatz
(\ref{eq:warpedvolmetric}) for $(p+1)$-dimensional dependent fluctuations of a $D$-dimensional scalar.

As discussed in Section \ref{sec:PertTheory}, the dilaton fluctuation transforms under 
$D$-dimensional diffeomorphisms as
\be
\delta \phi(x,y) \rightarrow \delta \phi + \xi^p \partial_p \phi_0(y).
\nonumber
\ee
Thus, there exists a gauge, ``unitary gauge," in which the dilaton fluctuation vanishes.  This can be
arrived at by making a gauge transformation with gauge parameter 
$\xi^p = -u(x)\partial^p \phi_0 \t \phi(y) /(\nabla \phi_0)^2$, where we have made the replacement
$\delta \phi(x) = u(x)$.  In this gauge all of the fluctuations
appear in the metric, which takes the (linearized) form:
\be
ds^2 &=& e^{2A_0(y)} e^{2\Omega_0(y)} \left\{ \hat g_{\mu\nu} 
	\left[1+u(x)(\delta A + \delta \Omega + \frac{\nabla \phi_0 \cdot \nabla A_0}{(\nabla \phi_0)^2} \t\phi(y))\right] + 2e^{(p-3)\Omega_0} \hat \nabla_\mu \partial_\nu u(x) E(y)\right\}\next
	&& + e^{-2\left(\frac{p+1}{D-p-3}\right) A_0(y)} 
		\left\{\t g_{mn}\left[1-u(x)\left(\frac{p+1}{D-p-3}\right)(\delta A-\frac{\nabla \phi_0 \cdot \nabla A_0}{(\nabla \phi_0)^2}\t \phi(y)\right] \right. \next
	&& \hspace{.8in} \left. + \t\nabla_{(m} \left(\frac{\partial_{n)}\phi_0}{(\nabla \phi_0)^2}\t \phi(y)\right) u(x)\right\} + \partial_\mu u(x) \t \phi(y) e^{-2A_0(y)} \frac{\partial_m \phi_0}{(\nabla \phi_0)^2} dx^\mu dy^m.
\ee
Notice that in this gauge the internal metric has off-diagonal terms, and the metric is no longer block-diagonal
in the internal and external directions.

In many cases of interest, the background dilaton profile $\phi_0(y)$ is related to the 
background warp factor $\phi_0(y) = q A_0(y)$, for some $q$.
For example, backreaction from p-branes create warping of precisely this form, as we will explore in more
detail in the next section.
The breathing mode then appears in the dilaton through the ansatz:
\be
\phi(x,y) = q A(y,u(x)),
\label{eq:phiProfile}
\ee
so that to linear order we have $\phi(x,y) \approx qA_0(y) + u(x) q \delta A + {\mathcal O}(u^2)$,
e.g.~$\t\phi(y) = q \delta A$.
The constraint equation (\ref{eq:dilatonConst})
can then be written in a simple form:
\be
-\left(\frac{D-2}{D-p-3}\right) \frac{e^{QA}}{Q} \partial_\mu \partial_m \left(e^{-QA(y,u(x))}\right) = 0
\ee
where $Q \equiv (p+1) + \frac{q^2(D-p-3)}{2(D-2)}$.  
Similarly as in \cite{FTUD}, the solution is a {\it shift ansatz} for the warp factor
\be
e^{-QA(y,u(x))} = e^{-QA_0(y)} + u(x).
\label{eq:GravityScalarShiftAnsatz}
\ee
The dilaton fluctuation does not contribute to the 
$\mu\neq \nu$ constraint equation (\ref{eq:GaugeInvConstraintmunu}) at linear order, which
taking advantage of the shift form of the warp factor, takes the form
\be
\t \nabla^2 E(y) = \frac{2\gamma}{Q} \left[\frac{e^{-2\gamma A(y,u(x))}}{\t V_{D-p-1}} 
	- \frac{e^{(Q-2\gamma)A(y,u(x))}}{\t V_{D-p-1}} \int\sqrt{\t g}\, e^{-2\gamma A(y,u(x))}\right]\, .
\label{eq:CompensatorGravityScalarGeneral}
\ee
Notice that the left hand side of this equation is independent of the breathing mode $u(x)$, while
the right hand side in general depends on $u(x)$.
However, if $Q= 2\gamma$ (\ref{eq:CompensatorGravityScalarGeneral}) becomes,
\be
\t \nabla^2 E(y) = e^{-QA(y,u(x))} - \frac{\int \sqrt{\t g}\, e^{-QA(y,u(x))}}{\t V_{D-p-1}} = e^{-2\gamma A_0(y)} 
	- \frac{\int\sqrt{\t g}\, e^{-2\gamma A_0(y)}}{\t V_{D-p-1}}\, ,
\ee
so that the right hand side only depends on the background warp factor $A_0(y)$ and both sides of the equation 
are now manifestly independent of the warped breathing mode $u(x)$.  We will take this value of $Q$ from
now on.

Summarizing, spacetime dependent fluctuations of the D-dimensional metric and dilaton of the form:
\be
ds^2 &=& e^{2A(y,u(x))} e^{2\Omega[u(x)]} \left[\hat g_{\mu\nu} + 2e^{(p-3)\Omega} \hat \nabla_\mu \partial_\nu u(x)\ E(y)\right] dx^\mu dx^\nu  \next
 &&+ e^{-2\left(\frac{p+1}{D-p-3}\right)A(y,u(x))} \t g_{mn}(y) dy^m dy^n\, ; \hspace{.3in}
 \label{eq:WarpedGravityScalarMetric} \\
\phi(x,y) &=& q A(y,u(x)) \ \ \mbox{    where    }\ \  q^2 = 4\frac{(D-2)^2}{(D-p-3)^2}-\frac{2(p+1)(D-2)}{D-p-3}; \label{eq:WarpedScalarAnsatz}\\
e^{-2\gamma A(y,u(x))} &=& e^{-2\gamma A_0(y)} + u(x) ; \label{eq:GravityScalarShiftAnsatzSummary}\\
e^{(p-1)\Omega[u(x)]} &=& \frac{\t V_{D-p-1}}{\int \sqrt{\t g}\, e^{-2 \gamma A(y,u(x))}} 
	= \frac{1}{u(x) + \t V_W^{(0)}/\t V_{D-p-1}};  \\
\t \nabla^2 E(y) &=& e^{-2\gamma A(y,u(x))}-e^{-(p-1)\Omega[u(x)]} 
	= e^{-2\gamma A_0(y)} - \t V_W^{(0)}/\t V_{D-p-1}\, ,
\label{eq:WarpedGravityScalarComp}
\ee
satisfy all constraint equations, and thus are a consistent ansatz for spacetime fluctuations.
The warped volume modulus and the dilaton have been forced by the constraint equations to combine into a 
single $(p+1)$-dimensional degree of freedom, the {\it warped breathing mode} $u(x)$.

In the weakly warped limit, $e^{-2\gamma A_0(y)}\approx 1$, we have $e^{(p-1)\Omega} \simeq (u(x)+1)^{-1}$ and 
from (\ref{eq:WarpedGravityScalarComp}) the compensator vanishes to leading order
$\t \nabla^2 E \simeq 0$ so $E(y) \approx 0$.
Taking $u(x)+1 = e^{2(D-2) \beta/(p+1)\,\varphi(x)}$ with $\beta$ as in Section \ref{sec:unwarped}, 
the metric (\ref{eq:WarpedGravityScalarMetric}) becomes
\be
ds^2 \simeq e^{2\alpha \varphi(x)} \hat g_{\mu\nu} dx^\mu dx^\nu + e^{2\beta \varphi(x)}\t g_{mn}dy^m dy^n
\label{eq:weaklywarpedmetric}
\ee
which is the metric for the unwarped volume modulus (\ref{eq:unwarpedmetric}).
In the completely unwarped limit $e^{-\gamma A_0(y)} = 1$ of
(\ref{eq:WarpedGravityScalarMetric}-\ref{eq:WarpedGravityScalarComp})
the unwarped volume modulus and the dilaton decouple, as they are no longer forced to be
related by the off-diagonal constraint equation.
In the language of constructing gauge-invariant variables, since the background profile for the dilaton
is constant in the completely unwarped limit, gauge transformations no longer mix the metric and 
dilaton degrees of freedom.
However, in the weakly warped limit the slowly varying background profile for the dilaton does
still spontaneously break the $D$-dimensional gauge invariance and the metric and dilaton degrees of freedom do
still mix non-dynamically, 
so we cannot in principle decouple these degrees of freedom.
In this sense the completely unwarped limit $e^{-\gamma A_0(y)} =1$ of
(\ref{eq:WarpedGravityScalarMetric}-\ref{eq:WarpedGravityScalarComp}) is a singular limit - it is not smoothly
connected to the weakly warped limit because the action of gauge transformations on the dilaton
is not smoothly connected to the unwarped limit.

In particular, write the background warp factor as $A_0(y) = \epsilon f(y)$, so that $\epsilon$ controls the
strength of the warping; in the large volume limit we expect $\epsilon$ to be inversely proportional to the 
volume.
Small $\epsilon$ is the weakly warped limit, since the warp factor is then approximately constant over
the internal space 
\be
e^{2A(x,y)} \approx 1 -\gamma^{-1} u(x) + 2\epsilon f(y).
\nonumber
\ee
Derivatives of the background warp factor in the internal direction are proportional to $\epsilon$ in this limit, 
$\partial_m A_0 \sim \partial_m \delta A = \epsilon \partial_m f(y)$.
The dilaton in the weakly warped limit is, from (\ref{eq:phiProfile})
\be
\phi(x,y) \approx q \epsilon f(y) - \frac{q}{2\gamma} u(x), 
\nonumber
\ee
so again derivatives of the background dilaton in the internal direction are proportional to $\epsilon$ in
this limit, $\partial_m \phi_0 \sim q \epsilon \partial_m f(y)$.
From (\ref{eq:dilatonConst}) we see that the dilaton and warped volume modulus are coupled through
derivatives of the background dilaton and warp factor profiles, $\partial_m A_0, \partial_m \delta A, \partial_m \phi_0$.
But since each of these terms scales with the same power of $\epsilon$, the coupling does not become 
parametrically small in the weakly warped, small $\epsilon$, limit. More precisely, the constraint 
equation becomes
\be
\left(\frac{D-2}{D-p-3}\right) \left[\epsilon \partial_m f(y) + \frac{(p+1)}{2\gamma} \epsilon \partial_m f(y)\right]
	= \frac{q^2}{4\gamma}\epsilon \partial_m f(y)\, .
\nonumber
\ee
The strength of the warping $\epsilon$ completely cancels from this constraint equation, so it
is independent of the size of the warping, as long as the warping is non-zero.
Thus, the dilaton and warped volume modulus combine into the single warped breathing mode for all 
finite values of the warping, even the weakly warped limit.
This has important implications for the construction of effective theories from flux compactifications,
where the weakly warped limit is commonly used and the dilaton and volume modulus are assumed to be
independent degrees of freedom.  We will discuss this more in Section \ref{sec:Discussion}.

\subsection{Dimensionally Reduced Kinetic Term}
\label{sec:kinetic}

The quadratic effective kinetic term for the warped breathing mode $u(x)$ for the system 
(\ref{eq:WarpedGravityScalarMetric}-\ref{eq:WarpedGravityScalarComp}) 
arises from the dimensional reduction of the $D$-dimensional Ricci scalar and dilaton action
(for notational convenience, we will denote determinants of metrics as $\mbox{det}\,\hat g_{\mu\nu} = \hat g$
and $\mbox{det}\,\t g_{mn} = \t g$):
\be
S_{eff}^{kin} &=& \frac{1}{2\kappa_D^2}\int \sqrt{g_D} \left[R_D-\frac{1}{2}(\partial \phi)^2\right]
	= -\frac{1}{4\kappa_D^2} \int \sqrt{\hat g}\sqrt{\t g} \left[\delta g^{MN}\delta G_{MN} + \partial_\mu \delta \phi \partial^\mu \delta\phi\right] \\
	&=&  -\frac{1}{4\kappa_D^2} \int \sqrt{\hat g}\sqrt{\t g} \left[\delta g^{mn}\delta G_{mn} + \partial_\mu \delta \phi \partial^\mu \delta\phi\right] = \int \sqrt{\hat g} \left({\mathcal G}_{uu}^{(g)}+{\mathcal G}_{uu}^{(\phi)}\right) (\partial_\mu u(x))(\partial^{\hat \mu}u(x)).\nonumber
\ee
In moving to the last line we used the fact that the kinetic contribution to the fluctuated external Einstein 
tensor $\delta G_{\mu\nu}$ vanishes once the constraints are satisfied.  

We will first focus on the gravity contributions to the effective kinetic term $G_{uu}^{(g)}$.
After using the equation for the wavefunction for $E(y)$ (\ref{eq:WarpedGravityScalarComp}),
the field space metric from gravity can be written as
\be
{\mathcal G}_{uu}^{(g)} = &&-\frac{1}{4\kappa_D^2} \int \sqrt{\t g}\, e^{(p-1)\Omega}\left(\frac{p+1}{D-2}\right)
	\left[\frac{D-2}{p-1}\, e^{(p-1)\Omega}+\frac{D-p-3}{2}\, e^{2\gamma A}\right.\next
	&& \left.-\frac{D-p-3}{2}\, e^{(p-1)\Omega} \partial^{\t m} e^{2\gamma A} \partial_m E\right].
\ee
The last term can be integrated by parts, which gives a term proportional $\t \nabla^2 E$.  Using  (\ref{eq:WarpedGravityScalarComp}) again the warp factor dependent pieces completely cancel
out, and the numerical coefficients of the different terms miraculously combine: 
\be
{\mathcal G}_{uu}^{(g)} 
= -\left(\frac{(p+1)^2(D-p-1)}{8\kappa_{p+1}^2(D-2)(p-1)}\right) \frac{1}{(u(x)+\t V^{(0)}_W/\t V_{D-p-1})^2}\, .
\label{eq:kineticGravity}
\ee
The coefficient of the kinetic term (\ref{eq:kineticGravity}) exactly matches that of the unwarped volume
modulus (\ref{eq:unwarpedshiftkinetic}), up to a constant shift $u \rightarrow u + \t V^{(0)}_W/\t V_{D-p-1}$.

The kinetic term coming from the dilaton is:
\be
{\mathcal G}_{uu}^{(\phi)} = -\frac{1}{4\kappa_D^2} \int \sqrt{\t g}\, e^{-2\gamma A} e^{(p-1)\Omega} q^2 (\delta A)^2 = -\frac{1}{4\kappa_D^2} \int \sqrt{\t g}\, e^{(p-1)\Omega} \frac{q^2}{4\gamma^2}\, e^{2\gamma A}\, .
\label{eq:kineticScalar}
\ee
Combining (\ref{eq:kineticGravity}) and (\ref{eq:kineticScalar}), we can write the entire effective kinetic term
as,
\be
S_{eff}^{kin} = &&-\frac{1}{2\kappa_{p+1}^2}\int \sqrt{\hat g} 
	\frac{\partial_\mu u(x) \partial^{\hat \mu} u(x)}{(u(x)+\t V^{(0)}_W/\t V_{D-p-1})^2} \times \\
	&& \left[\frac{(p+1)^2(D-p-1)}{4(D-2)(p-1)}+\frac{q^2(D-p-3)}{8(D-2)^2} 
		\int \sqrt{\t g}\, \t V_{D-p-1}^{-1} \frac{u(x)+\t V^{(0)}_W/\t V_{D-p-1}}{u(x)+e^{-2\gamma A_0(y)}}\right] \nonumber
\ee
In the weakly warped limit $ e^{-2\gamma A_0(y)}\approx 1$, this simplifies considerably to the form,
\be
S_{eff}^{kin} = -\frac{1}{2\kappa_{p+1}^2} \int \sqrt{\hat g}\ 
	\left(\frac{p}{p+1}\right) \frac{\partial_\mu u(x) \partial^{\hat \mu} u(x)}{(u(x)+\t V^{(0)}_W/\t V_{D-p-1})^2} \, .
\ee

\section{Breathing mode of compact p-brane solutions}
\label{sec:pbranes}
\setcounter{equation}{0}

As we saw in the previous section, when the background dilaton is related to the background warp factor in a particular
way (\ref{eq:phiProfile}) the constraints for the warped breathing mode simplify considerably.
One set of examples where this happens is when the background is sourced by p-branes.

We start with the effective action for $D$-dimensional gravity, a dilaton, and a $(p+2)$-form gauge field 
$F_{p+2} = d C_{p+1}$:
\be
S = \frac{1}{2\kappa_D^2} \int d^Dx \sqrt{g_D}\left[ R_D - \frac{1}{2}(\partial \phi)^2 - \frac{e^{-\lambda\phi}}{2 (p+2)!} F_{p+2}^2\right]+ S_{loc}\, ,
\label{eq:Seffpbrane}
\ee
where $S_{loc}$ denotes the action for localized sources charged under the $C_{p+1}$-field.
For the usual D-branes in 10-dimensions, $\lambda = (3-p)/2$, but in general we will only require
\be
\lambda^2 = 4 - \frac{2 (p+1)(D-p-3)}{(D-2)}\, .\nonumber
\ee
The effective action (\ref{eq:Seffpbrane}) has p-brane background solutions \cite{Townsend:1987yy,Horowitz:1991cd,Stainlesspbranes},
\be
\label{eq:pbranestaticmetric}
ds^2 &=& H_0(y)^{2a} \hat \eta_{\mu\nu} dx^\mu dx^\nu + H_0(y)^{2b} \t g_{mn}(y) dy^m dy^n; \\
\label{eq:pbranestaticdilaton}
e^{-\phi} &=& H_0(y)^{\lambda/2}; \\
C_{p+1} &=& \pm H_0(y)^{-1} \hat \epsilon_{p+1}\, ;
\label{eq:pbranesstaticpform}
\ee
where we have taken the internal space now to be Ricci flat $\t R_{mn}(\t g) = 0$.  The exponents in the metric
are defined as,
\be
a = -\frac{(D-p-3)}{2(D-2)},\ \ \ \ \ b = \frac{p+1}{2(D-2)},
\ee
and $H_0(y)$ is a harmonic function on the internal space, satisfying (for localized sources)
\be
\t \nabla^2 H_0(y) = \sum_n Q_n \delta^{(D-p-1)}(y-y_n)\, .
\label{eq:harmonicH}
\ee
This background can be generalized to include additional background fluxes as well.  These background fluxes
act like an effective $p$-brane charge, and generalize (\ref{eq:harmonicH}) to include flux contributions
on the right hand side as sources \cite{vanRietSmearing}.
For example, in $D=10$ type IIB supergravity \cite{GKP}, 
additional $3$-form fluxes $G_3$ can behave like an effective 
$D3$-brane charge and contribute additional terms to the right hand side of (\ref{eq:harmonicH}) as
$ \t \nabla^2 H_0 = \frac{|\tilde G_3|^2}{12 g_s} + \mbox{localized terms}$.
In the smeared limit, the background solutions
(\ref{eq:pbranestaticmetric}-\ref{eq:pbranesstaticpform}) are just T-dual to the 
GKP \cite{GKP} background. However, this background ansatz, and our warped breathing mode ansatz
below, are more general, since they also apply in the limit of localized sources as well, where the T-
duality rules do not apply.
The background (\ref{eq:pbranestaticmetric}-\ref{eq:pbranestaticdilaton}) is exactly of the form proposed
in the last section for the relation between the warp factor and dilaton backgrounds.

From (\ref{eq:harmonicH}) the equation for the harmonic function $H_0$ is unchanged by
the shift $H \rightarrow H + u$ where 
$u=u(x)$ is a constant on the internal space. 
This shift acts like a warped volume modulus on the metric (\ref{eq:pbranestaticmetric}), fluctuating the
warp factor.  However, it also appears as a fluctuation of the dilaton and $(p+1)$-form gauge potential $C_{p+1}$.
The fluctuation in $C_{p+1}$ will not play any role in determining the kinetic dynamics, e.g.~it does not appear
in the constraints, and will only affect the (flat) potential for $u$.  However, the fluctuation of the dilaton
is precisely the same as for the warped breathing mode (\ref{eq:WarpedScalarAnsatz}) from Section 
\ref{sec:BreathingMode}.
Thus, the warped breathing mode $u(x)$ from the previous section arises 
naturally from the shift\footnote{A similar shift was
also found in \cite{Maeda}, but with a restricted form of the ansatz that limits the physical interpretation.}
 invariance $H \rightarrow H + u$ of the static $p$-brane background (\ref{eq:harmonicH}).

Now let us consider a general spacetime-dependent deformation of the harmonic function 
$H_0(y) \rightarrow H(y,u(x))$, 
where the spacetime dependence in the harmonic function arises through the spacetime-dependent shift $u(x)$.  
In order for this
deformation to be a ``good" deformation it must satisfy all of the constraint equations.  In addition to the constraint
equations coming from the Einstein equations, which we have seen before, we must also satisfy constraint equations
coming from the $F_{p+2}$-form equations of motion, as well as be able to consistently 
solve the dilaton equation of motion.
In order to solve the all of the constraint equations, our ansatz for the metric must include 
a $u(x)$-dependent Weyl factor on the $(p+1)$-dimensional metric and a ``compensator" $E(y)$:
\be
ds^2 &= &H(y,u(x))^{2a} e^{2\Omega[u(x)]} \left[\hat \eta_{\mu\nu} + 2 e^{(p-3)\Omega}\hat \nabla_\mu \partial_\nu u(x) E(y)\right] dx^\mu dx^\nu \next
&&+ H(y,u(x))^{2b} \t g_{mn}(y) dy^m dy^n\, .
\label{eq:pbranesdynamicalMetric}
\ee
To make contact with the previous section, we can rewrite (\ref{eq:pbranesdynamicalMetric}) in terms
of the warp factor $A(y,u(x))$ 
through the relation $A(y,u(x)) = a \log H(y,u(x))$.
The Weyl factor is
\be
e^{(p-1)\Omega} = \frac{\t V_{D-p-1}}{\int \sqrt{\t g} H(y,u(x))} = \frac{1}{u(x) + \t V^{(0)}_W/\t V_{D-p-1}}
\ee
where $\t V^{(0)}_W = \int \sqrt{\t g} H_0(y)$ is the background warped volume.
The dilaton and $C_{p+1}$-form fields also gain spacetime dependence through the harmonic function and the 
Weyl factor in the following way:
\be
\label{eq:pbranesdynamicalMatter1}
e^{-\phi} &=& H(y,u(x))^{\lambda/2} \\
C_{p+1} &=& \pm H(y,u(x))^{-1}\,e^{(p+1)\Omega}\, \bar \epsilon_{p+1}\, ;
\label{eq:pbranesdynamicalMatter2}
\ee
where $\bar \epsilon_{p+1}$ is the epsilon-tensor constructed from 
$\bar g_{\mu\nu} = \hat \eta_{\mu\nu} + 2e^{(p-3)\Omega} \hat \nabla_\mu \partial_\nu u(x) E(y)$.

The dilaton and $F_{p+2}$-form field do not contribute to the $\mu \neq \nu$ constraint equation,
which reads
\be
\t \nabla^2 E(y) = H(y,u(x)) - e^{-(p-1)\Omega} = H_0(y) - \frac{\int \sqrt{\t g}\, H_0(y)}{\int \sqrt{\t g}}\,.
\label{eq:pbraneCompensator}
\ee
Clearly (\ref{eq:pbraneCompensator}) integrates to zero on both sides, and is manifestly independent
of the warped breathing mode $u(x)$.
As in the previous section, the dilaton does contribute to the $\mu m$ constraint equation, but the $F_{p+2}$-form
field does not; in fact, $F_{p+2}$ does not contain any spacetime derivatives due to the structure of $C_{p+1}$.
The $\mu m$ constraint equation then reads,
\be
\delta G_{\mu m}-\kappa_D^2 \delta T_{\mu m} &=& -\left(pa + b(D-p-2)\right)\partial_\mu \partial_m \log H
	+ (D-2)ab \partial_\mu \log H \partial_m \log H \next
	&& - \frac{\lambda^2}{8}\partial_\mu \log H \partial_m \log H
	= -\frac{1}{2} H^{-1} \partial_\mu \partial_m H = 0\, .
\label{eq:Hmixed}
\ee
As anticipated, solutions to (\ref{eq:Hmixed}) are the {\it shift} solutions,
\be
H(y,u(x)) = u(x) + H_0(y)\, .
\label{eq:Hshift}
\ee

However, it is not enough to just solve the Einstein constraint equations.
We need to show that (\ref{eq:pbranesdynamicalMetric}-\ref{eq:pbranesdynamicalMatter2}) 
are solutions to the full D-dimensional equations of motion, including the equations of motion coming
from the $F_{p+2}$-form and the dilaton.
The equation of motion for the $F_{p+2}$ form is:
\be
-d \left[ H^{-a(p+1)+b (D-p-1)-2 + \lambda^2/2} \left( \t \star_y d_y H(y,u(x))\right)\right] 
	= \sum_n Q_n \delta^{(D-p-1)}(y-y_n) \hat \epsilon_{p+1}\, ,
\ee
where we denoted an exterior derivative in the internal direction as $d_y$.
Notice that the exponent vanishes, so this simplifies to 
\be
d\left[(\t \star_y d_y H(y,u(x))) \right] = \sum_n Q_n \delta^{(D-p-1)}(y-y_n) \t \epsilon_{D-p-1}\,.
\label{eq:pformEOM}
\ee
When the exterior derivative is in the internal direction we just find the condition that $H(y,u(x))$ must be harmonic
on the internal space.
Taking the exterior derivative to be in the $(p+1)$-spacetime direction, (\ref{eq:pformEOM}) becomes
\be
d_x \left[\t \star_6 d_y H(y,u(x))\right] = 0
\ee
which is satisfied identically for the shift form (\ref{eq:Hshift}) of the harmonic function.
The dilaton and internal Einstein equations of motion simplify as well, reducing to
\be
\hat \Box H(y,u(x)) = 0\, .
\ee
With the shift form of the harmonic function (\ref{eq:Hshift}), 
this just reduces to the (linearized) equation of motion for $u(x)$:
\be
\hat \Box u(x) = 0\, ,
\ee
which indicates that the warped breathing mode is massless.

Fluctuations of the warped breathing mode do not induce fluctuations of the p-brane itself
at linear order, as can be seen by inspecting the kinetic terms of the DBI action, so
(\ref{eq:pbranesdynamicalMetric}-\ref{eq:pbranesdynamicalMatter2}) also solve the p-brane equation
of motion.
It is important to note that the massless warped breathing mode studied here is an independent
degree of freedom from the degrees of freedom controlling the position of the p-brane in the internal
space. It would be interesting to see how these degrees of freedom couple in the dimensionally
reduced effective action, and we leave this for future work.

In summary, we have shown that the Einstein-dilaton-p-form system (\ref{eq:Seffpbrane}) on the background
(\ref{eq:pbranestaticmetric}-\ref{eq:pbranesstaticpform}) has a $(p+1)$-dimensional 
``warped breathing mode" $u(x)$, realized non-trivially in the D-dimensional fields through
(\ref{eq:pbranesdynamicalMetric}-\ref{eq:pbranesdynamicalMatter2}).

\section{Discussion}
\label{sec:Discussion}
\setcounter{equation}{0}

We have argued in this paper that in generic $D$-dimensional warped compactifications to $(p+1)$-dimensions, the
fluctuations associated with the warped volume modulus and the dilaton (a $D$-dimensional scalar
field with a non-zero profile in the compact directions) combine into a single $(p+1)$-dimensional 
degree of freedom, which we have called
the warped breathing mode.  As discussed in Section \ref{sec:PertTheory}, these fluctuations combine in the
presence of non-trivial warping because of
two effects.  First, the warping breaks the $D$-dimensional diffeomorphism invariance so that the fluctuations
transform non-trivially under diffeomorphisms.  
The gauge-invariant dilaton fluctuation (\ref{eq:GaugeInvDPhi}) contains mixing between the metric
and dilaton fluctuations. Second, the warping also
leads to non-trivial constraint equations involving the fluctuations 
arising from the $D$-dimensional Einstein equations.
We explicitly illustrated this in Section \ref{sec:BreathingMode} by constructing
the $D$-dimensional wavefunction for the warped volume modulus and the dilaton, and showed that solving
the Einstein constraint equations forces these fluctuations to combine into a single degree of freedom.  In 
Section \ref{sec:pbranes} we showed that the warped breathing mode is the natural zero mode on
the warped backgrounds sourced by p-branes, indicating that it is indeed 
the correct low-energy degree of freedom in the presence of objects that source warping.

The mechanisms of mixing discussed here (spontaneous breaking of diffeomorphisms 
and non-trivial constraint equations) are not restricted to a bulk scalar field and volume modulus.
P-form gauge fields that obtain a non-zero background profile will lead to similar mixings between the p-form
and metric degrees of freedom (see \cite{GM}).
Further, gauge transformations associated with the p-form fields themselves
can also mix degrees of freedom in different sectors when the
p-form fluxes have non-trivial backgrounds.  For example, in type IIB supergravity, Chern-Simons couplings between
the $4$-form and $2$-form gauge potentials $C_4, C_2, B_2$ implies that $C_4$ transforms under gauge 
transformations of the $2$-form potentials when the $2$-forms have background $3$-form field strengths, e.g.:
\be
C_2 & \rightarrow & C_2 + d\zeta_1^C; \next
B_2 &\rightarrow & B_2 + d\zeta_1^B; \next
C_4 &\rightarrow & C_4 + \frac{1}{2} \zeta_1^C\wedge H_3^{(0)} + \frac{1}{2} \zeta_1^B\wedge F_3^{(0)}.
\nonumber
\ee
Generalizing the gauge-invariance and constraint equation arguments given here,
fluctuations in $C_4$ will mix with those of $B_2, C_2$; this was seen explicitly in \cite{FreyPolchinski,FTUD} for
the axion of $C_4$ in GKP \cite{GKP} backgrounds.  Similar mixing effects of p-form fluctuations 
will likely arise in the ``generalized BPS-like" backgrounds of \cite{vanRietSmearing} or \cite{Lust:2008zd}.
Thus, the approach given here of identifying the gauge-invariant combinations of fluctuations that are independent
under the constraint equations is a useful organizing scheme for understanding the structure of
effective theories arising from compactification in general.

The effects discussed in this paper all arise for non-trivial warping, where the background profiles are non-constant
$\partial_m A_0(y), \partial_0\phi_0(y) \neq 0$.  In the weakly warped limit the background profiles
approach a constant 
e.g.~$e^{\phi_0(y)},e^{2A_0(y)} \sim 1 + 2 \epsilon f(y)$ for some small $\epsilon$,
where $\epsilon$ is inversely proportional to some power of the volume.
Thus, it seems we can sidestep the subtleties associated with warping as long as we are willing to work
at a sufficiently large volume where there are no strongly warped regions.
This line of argument certainly works to remove the problems due to warping of wavefunction 
localization and integrating out KK modes discussed in the introduction.
In the large volume limit, the gravitational potential well generated by the warping disappears, so that
wavefunctions spread out over the entire internal space.  Likewise, KK modes become hierarchically more massive
than the zero mode in the large volume limit (see 
\cite{GM,WarpedSUSY,FreyMaharana} for more discussion of these effects).

But it is hard to see how the non-dynamical mixing from the diffeomorphisms and constraint equations can
be removed by a large volume limit: in the completely unwarped limit the mixing between the dilaton and the
volume modulus vanishes.  At strong warping, however,
the dilaton and volume modulus combine into a single degree of freedom.  
At the level of the equations, we found in Section \ref{sec:BreathingMode} that the constraint equation
for the warped volume modulus $u(x)$ and the dilaton $\delta \phi(x)$ takes the (schematic) form:
\be
\delta G_{\mu m} - \kappa_D^2 \delta T_{\mu m} = 0 \Rightarrow 
	(\partial_\mu u(x)) \partial_m A_0(y) \sim (\partial_\mu \delta \phi(x))\partial_m \phi_0(y) ;
\nonumber
\ee
where the background profiles scale in the same way in the weakly warped limit
$\partial_m A_0, \partial_m \phi_0 \sim \epsilon \partial_m f(y)$.  
The scaling with the strength of the warping $\epsilon$ cancels out but 
the non-zero derivative of the profile does not.  The only way to solve this equation for any finite strength
of the warping $\epsilon$, then, is if the dilaton and the warped volume modulus
combine into a single breathing mode, even in the weakly warped limit.

Note that we have not solved the full set of linearized equations for a general set of perturbations 
on the most general background, so additional independent degrees of freedom may be present.
In particular, there should be KK modes of the warped breathing mode, but it is not at all clear
what form such fluctuations will take.  We have not attempted to study such perturbations.
Instead, however, we have shown that the dilaton cannot be
taken to be independent from the warped volume modulus in a warped background, as is commonly
done.
We leave the study of more general perturbation ans\"atze to future work.

\section*{Acknowledgments}
We would like to thank R.~Brandenberger, A.~Castro, K.~Dasgupta, D.~George, D.~Marsh, L.~McAllister, 
G.~Shiu, and Y.~Wang for useful related discussions and comments, and would particularly like
to thank A.~Frey and T.~Van Riet for comments on an earlier version of the paper.
B.U.~is supported in part by NSERC, an IPP (Institute of Particle Physics, Canada) Postdoctoral Fellowship, and
by a Lorne Trottier Fellowship at McGill University.

\appendix

\section{Warped Volume Modulus}
\label{sec:WarpedVolMod}
\setcounter{equation}{0}

In unwarped backgrounds the volume modulus is easy to identify: it is just a simple rescaling of the internal metric
(together with a Weyl rescaling of the spacetime metric so that the lower dimensional spacetime is in Einstein frame).
In warped backgrounds, the definition of the warped volume modulus is not as simple \cite{GM, FTUD}.
In \cite{FTUD} the warped volume modulus was constructed for warped compactifications from $10$ to $4$
dimensions, and it was seen there that the volume modulus mixes with the warp factor and gives rise
to additional ``compensator" terms in the metric.  In this Appendix we generalize
the construction of \cite{FTUD} to warped compactifications with arbitrary numbers of dimensions, seeing again that the
warped volume modulus mixes with the warp factor and gives rise to metric compensators.
In the unwarped limit these mixings and compensators vanish, so that the fluctuation 
reverts back to a simple rescaling of the internal metric.

The mode we would like
to study is the warped generalization of the volume modulus.  
In order to differentiate the warped volume modulus from other possible deformation modes, we will
require that it satisfies a few simple properties: the fluctuation should correspond, in some gauge, to a pure trace
fluctuation $\delta \varphi_{mn} \sim \t g_{mn}$; it should correspond to a fluctuation in the ``warped volume"
$\t V_W^{(0)} = \int \sqrt{\t g} e^{(p-1)A_0-(D-p-1)B_0}$; it should satisfy all of the constraints; and it should reduce
to the unwarped volume modulus in the unwarped limit.
We will first review the unwarped volume modulus, then construct an ansatz for the warped volume modulus that
meets the above criteria.

We are assuming that there is some bulk matter with energy-momentum
tensor $T_{MN}$, such that the background metric is a solution to the
background Einstein equations $G_{MN}-\kappa_D^2 T_{MN} = 0$ for a maximally symmetric
spacetime metric $\hat g_{\mu\nu}$ and arbitrary internal space $\t g_{mn}$.
We are taking the background matter fields (including the dilaton) to be fixed with no fluctuations, so they 
are only important for sourcing the background, and
we will not need their detailed form.

\subsection{Review: Unwarped Volume Modulus}
\label{sec:unwarped}

Let us first start by reviewing the unwarped volume modulus, following the notation of \cite{RoestThesis}.
The unwarped metric corresponds to constant warp factors,
which we will set to unity by rescaling the $x^\mu,y^m$ coordinates:
\be
ds_D^2 = \hat g_{\mu\nu}(x) dx^\mu dx^\nu + \t g_{mn}(y)dy^m dy^n\, .
\label{eq:unwarpedBackground}
\ee
A fluctuation of the volume modulus $\varphi$ corresponds to a fluctuation of the overall scale of the internal
metric,
\be
ds_D^2 = \hat g_{\mu\nu}(x) dx^\mu dx^\nu + e^{2\beta\varphi(x)} \t g_{mn}(y)dy^m dy^n\, ,
\ee
as can be seen by the fact that the internal volume scales with $\varphi$ as 
$V_{D-p-1} = e^{-(p-1)\varphi} \int \sqrt{\t g}$.
However, in order to remain in $(p+1)$-dimensional Einstein frame after compactification, we must 
also include a modulus-dependent Weyl rescaling of the $(p+1)$-dimensional spacetime,
\be
ds_D^2 = e^{2\alpha \varphi(x)} \hat g_{\mu\nu}(x) dx^\mu dx^\nu 
	+ e^{2\beta\varphi(x)} \t g_{mn}(y)dy^m dy^n\, ,
\label{eq:unwarpedmetric}
\ee
where $\beta = -\frac{(p-1)}{D-p-1}\alpha$ ensures the Einstein frame condition.  
This also leads to the identification of the lower dimensional Newton's constant $\kD = \t V_{D-p-1} \kp$, where
$\t V_{D-p-1} = \int \sqrt{\t g_{D-p-1}}$ is the $(D-p-1)$-dimensional (unwarped) volume.
As we will see soon, if we further choose
\be
\alpha^2 = \frac{D-p-1}{2(D-2)(p-1)}
\label{eq:alpha}
\ee
then $\varphi$ will be a canonically normalized scalar field in the resulting $(p+1)$-dimensional effective
theory.
For small spacetime fluctuations of the volume modulus:
\be
\varphi(x) = \varphi_0 + \delta \varphi(x) + ...
\ee
the fluctuation (\ref{eq:unwarpedmetric}) in the notation of warped perturbation theory from Section \ref{sec:warpedperttheory} corresponds to the gauge-invariant metric fluctuations,
\be
\Phi_{mn}(x,y) &=& \beta \delta \varphi(x) \t g_{mn}(y)\,; \next
\Psi(x,y) &=& -\alpha \delta \varphi(x)\, .
\ee

The constraint equations $\delta G_{\mu\nu}|_{\mu\neq \nu}, \delta G_{\mu m}$ are satisfied
identically for the ansatz (\ref{eq:unwarpedmetric}), and the part of the internal Einstein equation proportional
to the kinetic term is:
\be
G_{mn} &=& - e^{2(\beta-\alpha)\varphi_0} \t g_{mn} \hat \Box \delta \varphi \frac{\alpha(D-2)}{D-p-1} +...
\ee
where by $...$ we mean that only the kinetic pieces are shown.
We can construct the $(p+1)$-dimensional (quadratic) effective kinetic term for the volume modulus $\varphi$ 
by reducing the Ricci scalar, which becomes:
\be
S_{eff} &=& \frac{1}{2\kD}\int \sqrt{g_D}\, R_D = \frac{1}{4\kD}\int \sqrt{g_D}\, \delta G_{MN} \delta g^{MN} \next
&=& \frac{1}{4\kD} \int \sqrt{g_D}\, \delta G_{mn} \delta g^{mn} 
	= -\int \sqrt{\hat g_{p+1}}\, \frac{(\partial_\mu \delta \varphi)(\partial^{\hat \mu}\delta \varphi)}{4\kp}
\ee
where in the last step we used the definition of $\alpha$ above (\ref{eq:alpha}) so that $\varphi$ is
canonically normalized (the extra factor of $(2\kappa_{p+1})^{-1}$ is a common convention).
Another convenient parameterization of the volume modulus is in terms of the ``breathing mode" 
$u(x) = e^{2(D-2)\beta/(p+1) \varphi(x)}$
for which the metric and effective kinetic term become:
\be
ds_D^2 &=& u(x)^{-\frac{(p+1)(D-p-1)}{(D-2)(p-1)}} \hat g_{\mu\nu}dx^\mu dx^\nu 
	+ u(x)^{\frac{p+1}{D-2}} \t g_{mn} dy^m dy^n ; \\
S_{eff} &=& -\int \sqrt{\hat g_{p+1}}\ \left(\frac{(p+1)^2(D-p-1)}{8 \kappa_{p+1}^2 (D-2)(p-1)}\right)
	\frac{(\partial_\mu u(x))(\partial^{\hat \mu} u(x))}{u(x)^2}\, .
\label{eq:unwarpedshiftkinetic}
\ee

\subsection{Warped Volume Modulus}
\label{eq:subwarpedVolMod}

As discussed at the beginning of this section, we would like to construct an ansatz for the warped volume
modulus on the warped background
\be
ds^2 = e^{2A_0(y)} \hat g_{\mu\nu}(x) dx^\mu dx^\nu + e^{-2B_0(y)} \t g_{mn}(y) dy^m dy^n\, ,
\label{eq:warpedvolbackground}
\ee
where we will take $B_0(y) = (p+1)/(D-p-3) A_0(y)$ as in the main text.
The relevant ``warped volume" appears in the dimensional reduction of the $D$-dimensional Ricci scalar
(we will suppress subscripts on metric determinants as 
$\mbox{det} \hat g_{\mu\nu} = \hat g,\ \mbox{det}\t g_{mn} = \t g$):
\be
\frac{1}{2\kappa_D^2}\int \sqrt{g_D}\, R_D \supset \frac{1}{2\kappa_D^2} \int \sqrt{\hat g}\, \hat R_{p+1} \int \sqrt{\t g}\, e^{-(p-1)A_0-(D-p-1)B_0} = \frac{1}{2\kappa_D^2} \int \sqrt{\hat g}\,\hat R_{p+1} \t V_W^{(0)}\hspace{.25in}
\label{eq:RicciRed}
\ee
where
\be
\t V_W^{(0)} \equiv \int \sqrt{\t g}\, e^{-(p-1)A_0 - (D-p-1)B_0} = \int \sqrt{\t g}\, e^{-2\gamma A_0(y)}
\ee
is the warped volume, with $\gamma \equiv (D-2)/(D-p-3)$.

The ansatz for the warped volume modulus we will use is\footnote{A qualitative argument for this form is given
in the beginning of Section \ref{sec:WarpedBreathingAnsatz}.}:
\be
ds^2 &=& e^{2A(y,u(x))} e^{2\Omega[u(x)]} \left[\hat g_{\mu\nu} + 2 e^{(p-3)\Omega[u(x)]}\hat \nabla_\mu \partial_\nu u(x) E(y)\right] dx^\mu dx^\nu \next
&& + e^{-2\left(\frac{p+1}{D-p-3}\right)A(y,u(x))} \t g_{mn}(y) dy^m dy^n ,
\label{eq:warpedvolmetricA}
\ee
where we have promoted the warp factor to be a function of the warped volume modulus so that
at linear order in $u(x)$
\be
A(y,u(x)) &\approx & A_0(y) + u(x) \delta A(y) + {\mathcal O}(u^2). \nonumber
\ee
We have also included a Weyl factor, defined as
\be
e^{(p-1)\Omega} = \frac{\int \sqrt{\t g}}{\int \sqrt{\t g}\ e^{(p-1)A} e^{-(D-p-1)B}} 
 = \frac{\t V_{D-p-1}}{\t V^W_{D-p-1}}
\ee
so that the dimensionally reduced Ricci scalar is in Einstein frame, where again to linear order
$\Omega[u(x)] \approx \Omega_0 + u(x) \delta \Omega + {\mathcal O}(u^2)$ with
$e^{(p-1)\Omega_0} \equiv \t V_{D-p-1}/\t V_W^{(0)}$.  This implies $\kappa_D^2 = \t V_{D-p-1} \kappa_{p+1}^2$,
as in the unwarped case.
For convenience, we will denote the unwarped metric with the compensator piece as
\be
\bar g_{\mu\nu} = \hat g_{\mu\nu} + 2 e^{(p-3)\Omega}\,\hat \nabla_\mu \partial_\nu u(x)  K(y)\, .
\ee

The constraint equations come from the off-diagonal parts of the $D$-dimensional Einstein equations, and are:
\be
\delta G_{\mu\nu}|_{\mu\neq \nu} = (\hat \nabla_\mu \partial_\nu u)\Big\{ e^{2A+2B+(p-1)\Omega} \Big[\t \nabla^2 E &+& (D-p-3) \partial_m B \partial^{\t m}E - (p+1)\partial_m A \partial^{\t m}E\Big] \next
\label{eq:munuConstraint}
  &-& (1-p) (\delta \Omega+\delta A) + (D-p-1)\delta B\Big\}=0; \hspace{.4in} \\
\label{eq:mumConstraint}
\delta G_{\mu m} = -p \partial_\mu \partial_m A + (D-p-2) \partial_\mu \partial_m B &-& (D-2) \partial_\mu B \partial_m A=0 \, .
\ee
In the absence of dilaton fluctuations, the $(\mu m)$ constraint equation (\ref{eq:mumConstraint}) becomes,
\be
\delta G_{\mu m} &=& -\left(\frac{D-2}{D-p-3}\right) \frac{e^{(p+1)A}}{(p+1)} \partial_\mu \partial_m \left(e^{-(p+1)A} \right) = 0\, ,
\label{eq:Gmum}
\ee
which is solved by the {\it generalized shift ansatz} for the warp factor,
\be
e^{-(p+1)A(y,u(x))} = e^{-(p+1)A_0(y)} + u(x)
\label{eq:GeneralShift}
\ee
where $A_0(y)$ is a background warp factor.
The Weyl factor becomes,
\be
e^{(p-1)\Omega} = \frac{\t V_{D-p-1}}{\int \sqrt{\t g}\ e^{-2\gamma A(x,y)}}\, .
\label{eq:WeylGravity}
\ee

The constraint equation (\ref{eq:munuConstraint}) is solved by:
\be
\label{eq:KDef}
\t \nabla^2 E &=& e^{-2\gamma A} \left[2\gamma \delta A + (1-p)\delta \Omega\right]e^{-(p-1)\Omega} \\
&=&  \frac{2\gamma}{p+1} \left[\frac{e^{-2\gamma A}}{\t V_{D-p-1}} \int \sqrt{\t g}\ e^{(-2\gamma+p+1)A}- \frac{e^{(-2\gamma+p+1)A}}{\t V_{D-p-1}}\int \sqrt{\t g}\, e^{-2\gamma A}\right]\, .
\label{eq:CaseACompensator}
\ee

Summarizing, we have shown that the ansatz,
\be
ds^2 =&&\left( e^{-(p+1)A_0(y)}+u(x)\right)^{-2/(p+1)} e^{2\Omega[u(x)]} \left[\hat g_{\mu\nu} 
	+ 2 \hat\nabla_\mu \partial_\nu u(x) E(y)\right]dx^\mu dx^\nu \next
	&&+ \left(e^{-(p+1)A_0(y)}+u(x)\right)^{2/(D-p-3)} \t g_{mn} dy^m dy^n
\label{eq:warpedvolmetricsummary}
\ee
with $E(y)$ solving (\ref{eq:CaseACompensator}) and $\Omega$ defined by
(\ref{eq:WeylGravity}), solves the linearized warped constraint equations.  It should
be straightforward to generalize this to the non-linear level as in \cite{FTUD}. 
Clearly, if we turn off the volume modulus
fluctuation $u(x) = 0$, we return to the background (\ref{eq:warpedvolbackground}).
In the unwarped limit, $e^{(p-1)A_0(y)} \rightarrow 1$, with the identification $u(x)+1 = e^{(D-p-3) \beta \varphi(x)}$
the metric (\ref{eq:warpedvolmetricsummary}) becomes the ansatz for the unwarped
volume modulus (\ref{eq:unwarpedmetric}) as in Section \ref{sec:unwarped}.
This ansatz for the metric fluctuations meets our criteria for the warped volume modulus outlined
at the beginning of this section:
the wavefunction has a pure trace component, it corresponds to a fluctuation of the warped volume $\t V_W$,
it solves the constraint equations, 
and it reduces to the unwarped volume modulus in the unwarped limit.

\subsubsection{Kinetic Term}

The effective kinetic term for $u(x)$ can be obtained from a dimensional reduction of the $D$-dimensional
Ricci scalar:
\be
S_{eff} &=& \frac{1}{2\kappa_D^2} \int \sqrt{g_D}\, R_D = -\frac{1}{4\kappa_D^2}\int \sqrt{\hat g} \sqrt{\t g}\ \delta g^{MN} \delta G_{MN} \next
& =&  -\frac{1}{4\kappa_D^2}\int \sqrt{\hat g} \sqrt{\t g}\, \delta g^{mn} \delta G_{mn}\,.
\ee
In the last step we used the fact that the kinetic contribution to the fluctuated external Einstein 
tensor $\delta G_{\mu\nu}$ vanishes once the constraints are satisfied.
The kinetic part of the Einstein equation in the internal directions is
\be
\delta G_{mn} &=& \hat \Box u(x) \Big[\t g_{mn} e^{-2A-2B-2\Omega} \left\{ p(\delta A+\delta \Omega)-(D-p-2)\delta B\right\} \\
&& +e^{(p-3)\Omega} \left[ - \t \nabla_m \partial_n K - \left(\partial_m (A+B)\partial_n E + \partial_m E \partial_n(A+B)-\t g_{mn} \partial_p E \partial^{\t p}B\right) \right. \next
&& +\t g_{mn} \t \nabla^2 E+\partial_p A\partial^{\t p}E \t g_{mn} + 2\left\{(D-p-3)\partial_p B \partial^{\t p}E - (p+1)\partial_p A \partial^{\t p}E\right\}\big]\Big] +...\nonumber
\ee
The fluctuated internal metric is:
\be
\delta g_{mn} = -2 \delta B\ g_{mn}, \ \ \ \ \delta g^{mn} = g^{mp} g^{nq} \delta g_{pq} = -2 \delta B\ g^{mn} = -2 \delta B\ e^{2B} \t g^{mn}\, .
\ee
Using (\ref{eq:KDef}) we can write $\delta g^{mn} \delta G_{mn}$ as:
\be
\delta g^{mn} \delta G_{mn} &&= -2 u\hat \Box u \delta B e^{-2A-2\Omega} \left[ (D-2)(\delta \Omega+\delta A) +\gamma (D-p-3) e^{2\gamma A+2\Omega} \partial^{\t m}A\partial_m E \right].\hspace{.4in}
\ee
Writing the effective kinetic term as:
\be
-\frac{1}{4\kappa_D^2} \int \sqrt{g_D}\ \delta g^{mn} \delta G_{mn} = \int \sqrt{\hat g}\ G_{uu} u\hat \Box u,
\ee
the field space metric is:
\be
G_{uu} = \frac{1}{4\kappa_D^2} \int \sqrt{\t g} \frac{2(D-2)}{D-p-3} e^{(p-1)\Omega} e^{(p+1-2\gamma)A}
\left[\delta \Omega+\delta A + e^{2\gamma A+2\Omega} \partial^{\t m}A\partial_m E\right]\, .
\ee
The last term can be integrated by parts, leaving us with
\be
G_{uu} = -\frac{1}{2\kappa_{p+1}^2}\frac{D-2}{(p+1)(D-p-3)}\left[\frac{4p\gamma}{(p-1)(p+1)} e^{2(p-1)\Omega} \frac{\left(\t V^{(p+1-2\gamma)}_W\right)^2}{\t V_{D-p-1}} \right. \next
\left.+ \left(1-\frac{2\gamma}{p+1}\right)e^{(p-1)\Omega} \frac{V^{(2(p+1)-2\gamma)}_W}{\t V_{D-p-1}}\right],
\label{eq:Guu}
\ee
where we are using the notation $\t V_W^\alpha \equiv \int \sqrt{\t g} e^{\alpha A_0(y)}$.

To make contact with the unwarped case it is also convenient to change variables to
$u(x) = e^{-(D-p-3)\beta \varphi(x)}$, with corresponding effective kinetic term,
\be
S_{eff} &=& \int \sqrt{\hat g}\ G_{\varphi \varphi}\ \varphi \hat\Box \varphi \, ; \\
G_{\varphi \varphi} &=& -\frac{1}{2\kappa_{p+1}^2} \frac{(p-1)}{(D-p-1)(p+1)^2} e^{-2(D-p-3)\beta \varphi}
\left[\frac{4p(D-2)}{(p-1)} e^{2(p-1)\Omega} \frac{\left(\t V^{(p+1-2\gamma)}_W\right)^2}{\t V_{D-p-1}} \right. \next
&& \left.+ \left(D(p-1)-p^2-4p+1\right)e^{(p-1)\Omega} \frac{V^{(2(p+1)-2\gamma)}_W}{\t V_{D-p-1}}\right]\, .
\label{eq:Gvarphi}
\ee
In the unwarped limit, the volume factors all cancel out, and remarkably all of the factors of $D$ and $p$ 
in (\ref{eq:Gvarphi}) cancel as well,
leading to the canonically normalized kinetic term as in the previous section.  However, for non-trivial
warping we see that now the kinetic term for the ``traditional" volume modulus $\varphi$
is no longer canonically normalized in general.  Of course, it is straightforward to canonically normalize
the kinetic term by an appropriate field redefinition.  The point here is that the warped volume modulus
does not automatically have the same kinetic term as the unwarped volume modulus - they are
related by a field redefinition.
This is in contrast to that found in \cite{FTUD}, where the kinetic terms for the warped and unwarped
volume moduli were found to be identical, a fact crucial for verifying the conjectured ${\mathcal N}=1$ K\"ahler
potential for the volume modulus.  

\subsection{Special Cases}

An interesting set of special cases of the warped volume modulus emerges when $p+1 = 2\gamma$.
In these cases, the second term in
(\ref{eq:Guu}) vanishes, and the first term simplifies considerably, so that the field space metric becomes:
\be
G_{uu} = -\frac{1}{2\kappa_{p+1}^2} \left(\frac{p}{p+1}\right) \frac{1}{\left(u(x)+\t V_W^{(0)}/\t V\right)^2} .
\ee
Taking the maximal spacetime dimension to be $D= 11$,
the only integer values of $D$ and $p$ for which this is satisfied are:
\be
\begin{array}{l l l}
p=2, & D=11: &  D = 11 \rightarrow 2+1 \ (M2)\\
p=3, & D = 10: & D=10 \rightarrow 3+1 \ (D3)\\
p=5, & D = 10: & D=11 \rightarrow 5+1 \ (M5).\\
\end{array}\nonumber
\ee
These special cases are precisely those for which the warped product structure
of the volume modulus deformed metric (\ref{eq:warpedvolmetricA}) matches that of the 
corresponding $p$-brane solutions without a dilaton: $D3$-branes in $10$-dimensional supergravity,
and, more trivially, $M2$- and $M5$-branes in $11$-dimensional supergravity.

In particular, (\ref{eq:warpedvolmetricA}) for 
the case of $D=11$ reduced to $p+1 = 3$-dimensional Minkowski space has the 
metric (relabeling $2A(y,u(x)) = -\phi(y,u(x))$ by convention):
\be
ds_{11}^2 &=& e^{-\phi(y,u(x))} e^{2\Omega(u(x))} \left[\hat \eta_{\mu\nu} + 2e^{-\Omega} \partial_\mu \partial_\nu u(x) K(y)\right] dx^\mu dx^\nu \next
	&&+ e^{\frac{1}{2}\phi(y,u(x))} \t g_{mn}(y) dy^m dy^n
\label{eq:11to3Metric}
\ee
with the shift ansatz (\ref{eq:GeneralShift}) taking the form,
\be
e^{\frac{3}{2}\phi(y,u(x))} = e^{\frac{3}{2}\phi_0(y)} + u(x)\, .
\label{eq:11to3Shift}
\ee
The warp factor structure in (\ref{eq:11to3Metric}) is exactly the same as the supergravity solution corresponding
to an extremal M2-brane \cite{Duff:1990xz}.  Upon supersymmetric compactification to
$(2+1)$-dimensional Minkowski space on spacetime filling $M2$-branes,
supersymmetry demands in this case that the $8$-dimensional
compact space be a Calabi-Yau 4-fold \cite{hep-th/9605053,hep-th/9908088,hep-th/0004103}.
The warp factor in this background satisfies the equation (coming from the equation of motion for the $3$-form
potential)
\be
\t \nabla^2 e^{3\phi/2} = \t \star_8 \left(X_8 - \frac{1}{2} G\wedge G\right) - \sum_j \delta^8(y-y_j)\, ,
\label{eq:11to3WarpEq}
\ee
where we must include (self-dual) $4$-form flux $G$, and a topological term 
$X_8(R) =$\linebreak $\frac{1}{8\cdot 4!} \left(\mbox{tr}R^4 - \frac{1}{4}\left(\mbox{tr}R^2\right)^2\right)$
to cancel the $C_3$ tadpole of the $M2$-branes at the locations $y_i$,
\be
\frac{\chi_{CY_4}}{24} = Q_2 + \frac{1}{2} \int_{CY_4} G\wedge G\, .
\ee
We see that the shift ansatz for the volume modulus dependence in the warp factor 
(\ref{eq:11to3Shift}) is quite natural since it is a zero mode
of (\ref{eq:11to3WarpEq}), even though it was derived in a very different way.

The case of $D=10$ reduced to $p+1=4$-dimensional Minkowski space has the metric
(\ref{eq:warpedvolmetricA}) (renaming $u(x) = c(x)$):
\be
ds_{10}^2 = e^{2A(y,c(x))} e^{2\Omega[c(x)]} \left[\hat \eta_{\mu\nu} + 2 \partial_\mu \partial_\nu c(x) K(y)\right] 
dx^\mu dx^\nu
	+ e^{-2A(y,c(x))} \t g_{mn}(y) dy^m dy^n\, ,\hspace{.2in}
\label{eq:10to4Metric}
\ee
with the shift ansatz (\ref{eq:GeneralShift}) taking the form,
\be
e^{-4A(y,u(x))} = e^{-4A_0(y)} + c(x)\, .
\label{eq:10to4Shift}
\ee
Again, the warped product structure of the metric (\ref{eq:10to4Metric}) is identical to that of the corresponding
extremal $D3$-brane solution \cite{Horowitz:1991cd}.  Upon supersymmetric compactification to 
$(3+1)$-dimensional Minkowski space on $D3$-branes, again the warped product structure of the metric also 
takes this form, and supersymmetry demands in this case that the $6$-dimensional
compact space be a Calabi-Yau 3-fold \cite{hep-th/9908088,GKP}.
The warp factor in this background satisfies the equation (coming from the equation of motion for the RR 4-form)
\cite{GKP}:
\be
-\t \nabla^2 e^{-4A} = \frac{|\widetilde G_3|^2}{12\mbox{Im}\tau} + 2 \kappa_{10}^2 T_3 \rho_3^{\mbox{loc}},
\label{eq:10to4WarpEq}
\ee
where we must include (imaginary self-dual) $G_3$ flux and orientifold $O3$-planes to cancel
the tadpole from the $D3$-brane charges,
\be
\int_{{\mathcal M}_6} H_3 \wedge F_3 + Q_3^{\mbox{loc}} = 0\,.
\ee
Again, we see that the shift form of the volume modulus dependence of the warp factor
(\ref{eq:10to4Shift}) is a natural zero mode of the background (\ref{eq:10to4WarpEq}).
The metric (\ref{eq:10to4Metric}) is just the volume modulus deformation of the GKP background found
in \cite{FTUD}.  

The final ``special" case of $D=11$ reduced to $p+1=6$-dimensional Minkowksi space has the metric (\ref{eq:warpedvolmetricA}),
\be
ds_{11}^2 = e^{2A(y,u(x))} e^{2\Omega(u(x))} \left[\hat \eta_{\mu\nu} + 2 e^{2\Omega}\partial_\mu \partial_\nu u(x) K(y)\right]
dx^\mu dx^\nu
	+ e^{-4 A(y,u(x))} \t g_{mn}(y) dy^m dy^n\, , \hspace{.3in}
\label{eq:11to6Metric}
\ee
with the shift ansatz (\ref{eq:GeneralShift}) taking the form
\be
e^{-6A(y,u(x))} = e^{-6A_0(y)} + u(x)\, .
\label{eq:11to6Shift}
\ee
As expected, the warped product structure of (\ref{eq:11to6Metric}) is identical to the supergravity
solution of the extremal $M5$-brane \cite{Guven}.  Upon compactification to $(5+1)$-dimensional Minkowski
space on spacetime filling $M5$-branes, the warped product structure also takes this form \cite{Witten:1996mz},
while low energy supersymmetry requires the internal space to be the orientifolds\footnote{Strictly speaking
an orientifold of M-theory is not well-defined, since there is no worldsheet theory to construct an orientifold
with respect to.  In practice, orientifolds in M-theory are defined by lifting the orientifold action in $IIA$
to the M-theory fields, see \cite{Dasgupta:1995zm,Witten:1995em} for some discussion.} $T^5/{\mathbb Z}_2$ or
$K3\times S^1/{\mathbb Z}_2$ (and their orbifolds) \cite{Dasgupta:1995zm,Witten:1995em}.
The warp factor satisfies the equation (coming from the Bianchi identity for the 4-form field strength),
\be
-\tilde \nabla^2 e^{-6A} = \sum_i q_i \delta^{5}(y-y_i) \, ;
\label{eq:11to6WarpEq}
\ee
for $M5$-branes $q_i=1$.  In order to satisfy the tadpole constraint coming from (\ref{eq:11to6WarpEq})
we must have additional sources of negative $M5$-brane charge; in contrast to the $M2$-brane case,
flux cannot carry $M5$-brane charge, and so cannot be used to cancel this tadpole.  Fortunately, as discussed
in \cite{Witten:1995em}, the twisted sector fields at the fixed points of the orbifold action of the orientifold 
carry negative charge $q_i=-1/2$ (in just such a way that they are free of gravitational anomalies), 
so the tadpole condition coming from (\ref{eq:11to6WarpEq}) can be satisfied.
Again, we see that the shift form of the volume modulus in the warp factor
(\ref{eq:11to6Shift}) is a natural zero mode of the background (\ref{eq:11to6WarpEq}).

It is perhaps remarkable that our simple analysis of a warped volume modulus, without any explicit reference
to the form of additional matter, has led quite naturally to the $D3$-, $M2$-, and $M5$-brane backgrounds.

\bibliographystyle{utcaps2}
\bibliography{warpedgeneral}

\end{document}